\documentclass[fleqn,usenatbib]{mnras}
\usepackage{newtxtext,newtxmath}
\usepackage[T1]{fontenc}
\DeclareRobustCommand{\VAN}[3]{#2}
\let\VANthebibliography\thebibliography
\def\thebibliography{\DeclareRobustCommand{\VAN}[3]{##3}\VANthebibliography}
\usepackage{graphicx}	
\usepackage{amsmath}	
\usepackage{threeparttable}
\usepackage{multirow}
\usepackage{footnotebackref}

\newcommand{\aum}{AU\,Mic}

\newcommand{\pr}{P$_{\rm{rot}}$}
\newcommand{\vs}{$v \sin i_{\rm{rot}}$}
\newcommand{\kms}{km\,s$^{-1}$}
\newcommand{\ms}{m\,s$^{-1}$}
\newcommand{\mass}{M$_{\rm{S}}$}
\newcommand{\msun}{M$_{\odot}$}
\newcommand{\rsun}{R$_{\odot}$}
\newcommand{\rpb}{R$_{\rm{p,b}}$}
\newcommand{\porbb}{P$_{\rm{orb,b}}$}
\newcommand{\ttrb}{T$_{0,\rm{b}}$}
\newcommand{\rhopb}{$\rho_{\rm{p,b}}$}
\newcommand{\mpb}{M$_{\rm{p,b}}$}
\newcommand{\apb}{a$_{\rm{p,b}}$}
\newcommand{\kpb}{K$_{\rm{p,b}}$}
\newcommand{\rpc}{R$_{\rm{p,c}}$}
\newcommand{\apc}{a$_{\rm{p,c}}$}
\newcommand{\porbc}{P$_{\rm{orb,c}}$}
\newcommand{\ttrc}{T$_{0,\rm{c}}$}
\newcommand{\rhopc}{$\rho_{\rm{p,c}}$}
\newcommand{\mpc}{M$_{\rm{p,c}}$}
\newcommand{\kpc}{K$_{\rm{p,c}}$}
\newcommand{\mearth}{M$_{\rm \oplus}$}
\newcommand{\rearth}{R$_{\rm \oplus}$}
\newcommand{\rs}{R$_{\rm{S}}$}
\newcommand{\teff}{T$_{\rm{eff}}$}
\newcommand{\lgg}{$\log g$}

\newcommand{\lsun}{L$_{\odot}$}
\newcommand{\lstar}{L$_{\rm{S}}$}

\newcommand{\idi}{i$_{\rm{DI}}$}
\newcommand{\chisqr}{$\chi^{2}$}
\newcommand{\chisqrr}{$\chi_{\rm{r}}^{2}$}
\newcommand{\oeq}{$\Omega_{\rm{eq}}$}
\newcommand{\dome}{d$\Omega$}
\newcommand{\radd}{rad\,d$^{-1}$}

\defcitealias{plavchan2020}{P20}
\defcitealias{plavchan2009}{P09}
\defcitealias{martioli2021}{M21}
\defcitealias{white2019}{W19}
\defcitealias{gaia2018}{G18}
\defcitealias{afram2019}{A19}
\defcitealias{mamajek2014}{M14}
\defcitealias{Barragan2019}{Ba19}
\defcitealias{bossini2019}{Bo19}
\defcitealias{brandt2015}{Br15}
\defcitealias{gaspar2009}{G09}
\defcitealias{klein2021}{Kl21}
\defcitealias{kavanagh2021}{Ka21}
\defcitealias{fischer2019}{F19}
\defcitealias{zicher2022}{paper I}

\title[HARPS AU Mic activity]{One year of AU Mic with HARPS: II - stellar activity and star-planet interaction}

\author[B. Klein et al.]{
Klein Baptiste,$^{1}$\thanks{E-mail: baptiste.klein@physics.ox.ac.uk}
Norbert Zicher,$^{1}$
Robert D. Kavanagh,$^{2}$
Louise D. Nielsen,$^{1,3}$
\newauthor
Suzanne Aigrain,$^{1}$
Aline A. Vidotto,$^{2}$
Oscar Barrag\'an,$^{1}$
Antoine Strugarek,$^{4}$
Belinda Nicholson,$^{1,5}$
\newauthor
Jean-François Donati,$^{6}$
Jérôme Bouvier$^{7}$
\\
$^{1}$Department of Physics, University of Oxford, OX13RH, Oxford, UK \\
$^{2}$ Leiden Observatory, Leiden University, PO Box 9513, 2300 RA Leiden, The Netherlands \\
$^{3}$ European Southern Observatory, Karl-Schwarzschild-Straße 2, 85748 Garching bei München, Germany \\
$^{4}$ AIM, CEA, CNRS, Université Paris-Saclay, Université Paris Diderot, Sorbonne Paris Cité, F-91191 Gif-sur-Yvette, France \\
$^{5}$ University of Southern Queensland, Centre for Astrophysics, Toowoomba, Australia \\
$^{6}$Univ. de Toulouse, CNRS, IRAP, 14 av. Belin, 31400 Toulouse, France \\
$^{7}$ Univ. Grenoble Alpes, CNRS, IPAG, 38000 Grenoble, France
}

\date{Accepted. Received in original form 2022 March 15}

\pubyear{2022}

\begin{document}
\label{firstpage}
\pagerange{\pageref{firstpage}--\pageref{lastpage}}
\maketitle

\begin{abstract}
We present a spectroscopic analysis of a 1-year intensive monitoring campaign of the 22-Myr old planet-hosting M dwarf \aum\ using the HARPS spectrograph. In a companion paper, we reported detections of the planet radial velocity (RV) signatures of the two close-in transiting planets of the system, with respective semi-amplitudes of 5.8\,$\pm$\,2.5\,\ms\ and 8.5\,$\pm$\,2.5\,\ms\ for \aum\,b and \aum\,c. Here, we perform an independent measurement of the RV semi-amplitude of \aum\,c using Doppler imaging to simultaneously model the activity-induced distortions and the planet-induced shifts in the line profiles. The resulting semi-amplitude of 13.3\,$\pm$\,4.1\,\ms\ for \aum\ c reinforces the idea that the planet features a surprisingly large inner density, in tension with current standard models of core accretion. Our brightness maps feature significantly higher spot coverage and lower level of differential rotation than the brightness maps obtained in late 2019 with the SPIRou spectropolarimeter, suggesting that the stellar magnetic activity has evolved dramatically over a $\sim$1-yr time span. Additionally, we report a 3-$\sigma$ detection of a modulation at 8.33\,$\pm$\,0.04\,d of the He\,I D3 (5875.62\,{\AA}) emission flux, close to the 8.46-d orbital period of \aum\,b. The power of this emission (a few 10$^{17}$\,W) is consistent with 3D magnetohydrodynamical simulations of the interaction between stellar wind and the close-in planet if the latter hosts a magnetic field of $\sim$10\,G. Spectropolarimetric observations of the star are needed to firmly elucidate the origin of the observed chromospheric variability.
\end{abstract}





\section{Introduction}\label{sec:section1}

Studying the processes driving planet formation and evolution is key to understand the large diversity in the distribution of exoplanets observed so far \citep{raymond2020}. This motivates efforts to detect and characterise planets around pre-main-sequence (PMS) stars. Close-in transiting planets are key targets in this venture. Their bulk density, accessible by measuring both the planet masses, by monitoring the radial velocity (RV) of their host star, and radii, by measuring the depth of their photometric transit curve, is a critical parameter to refine planet formation and evolution models \citep[e.g.,][]{mordasini2012,owen2017}. On the other hand, their atmospheric composition, accessible via transmission spectroscopy, and orbital parameters (e.g., orbit eccentricity and spin-orbit angle) enclose crucial information about their formation \citep[see the reviews of][]{baruteau2016,madhusudhan2019}. Finally, signatures of star-planet interactions such as anomalous stellar activity and flares can, in principle, yield constraints on the planet magnetic properties \citep{cuntz2000,saar2001,fischer2019}.

PMS stars are faint in the optical, and exhibit intense magnetic activity whose manifestations, such as large spots or frequent flares, induce photometric and radial velocity (RV) signals which hamper the search for close-in planets around them \citep[e.g.,][]{bouvier1989,berdyugina2005,crockett2012}. As a consequence, only a handful young ($<$\,100\,Myr) close-in planetary systems have been revealed so far: most of them are transiting planets detected by the \textit{K2} and \textit{TESS} space missions \citep{David2016a,Mann2016a,Mann2016b,David2019b,plavchan2020,rizzuto2020,mann2021}, and the remainder are non-transiting planets unveiled from the RV wobbles of their host star \citep{donati2016,yu2017}. So far, only two systems younger than 100\,Myr, namely \aum\ and V1298\,Tau, have planets with well-constrained mean bulk density measured \citep{klein2021,cale2021,suarez2021}.

Located in the 22-Myr $\beta$\,Pic moving group \citep{mamajek2014,malo2014,messina2016}, AU Microscopii (\aum, GJ\,803) is the second closest PMS star \citep[D\,=\,9.7248\,pc;][]{gaia2018}. This system has been intensively monitored over the past 50 years for its intense magnetic activity notably inducing $\sim$\,0.1-mag quasi-periodic photometric variations \citep[e.g.,][]{torres1973,rodono1986,plavchan2020,martioli2021}, and frequent flaring events of variable intensity \citep[e.g.,][]{robinson2001,hebb2007,macgregor2020,martioli2021}. The detection of two transiting Neptune-sized planets from Spitzer and TESS light curves \citep[TESS sectors 1 and 27;][]{plavchan2020,martioli2021} has increased the already high interest of the scientific community in the system which has been intensively monitored with various state-of-the-art spectrometers since then. These follow-up observations made it possible to measure a zero-compatible spin-orbit angle \citep{addison2020,hirano2020,martioli2020,palle2020}, and a bulk density of 1.32\,$\pm$\,0.2\,g\,cm$^{-3}$ for the close-in planet \aum\,b \citep{klein2021,cale2021}. Additionally, a 5$\sigma$-upper limit of 20.13\,\mearth\ has been reported for \aum\,c in \citet{cale2021}.


We observed the \aum\ system intensively with HARPS over a 10-month period, with the aim of precisely-capturing the RV signals induced by the stellar magnetic activity in order to better remove it and detect the two transiting planets \citep[][hereafter \citetalias{zicher2022}]{zicher2022}. From a total of 82 observations, we reported 2.4- and 3.4-$\sigma$ detections of \aum\,b and \aum\,c, with estimated masses of 11.7$^{+5.0}_{-4.9}$\,\mearth\ and 22.2$_{-6.6}^{+6.7}$\,\mearth, respectively. These masses significantly differ from the most recent estimates of \citet[][20.1\,$\pm$\,1.6\,\mearth\ and 5$\sigma$-upper limit of 20.13\,\mearth\ for the masses of \aum\,b and c, respectively]{cale2021}, and appear in tension with standard core-accretion models where planet c, about twice as massive as planet b, seems to have accreted less H/He material than the latter \citepalias[see Sec.~4.5 of][]{zicher2022}.

In the present study, we analyse the magnetic activity of \aum\ from the HARPS spectra published in \citetalias{zicher2022}. The intense magnetic field of the star and the small orbital distances of the transiting planets make this system an excellent candidate to search for planet-induced chromospheric emissions \citep[][Strugarek et al., 2022, submitted]{strugarek2019}. The HARPS spectral range allows us to access numerous chromospheric emission lines and search for potential modulations that could be explained by star-planet interactions \citep[][]{cuntz2000,saar2001,fares2010,cauley2019,strugarek2019}. On the other hand, the dense sampling of the stellar rotation cycle makes this data set ideally-suited to constrain the brightness distribution at the stellar surface using Doppler imaging (DI), while searching for planet-induced signatures shifting the line profiles \citep[see][]{petit2015,donati2016,yu2017,klein2021}. After describing the observations and data reduction in Sec.~\ref{sec:section2}, we investigate the chromospheric activity of the star and search for potential planet-induced modulations in Sec.~\ref{sec:section3}. In Sec.~\ref{sec:section4}, we use DI to invert the time series of HARPS cross-correlation functions (CCFs) while simultaneously searching for the RV signatures of the two close-in planets. We finally conclude, in Section~\ref{sec:section5}, by comparing our results to the literature and presenting future prospects for the system.

\section{Observations and data reduction}\label{sec:section2}

\begin{table*}
    \centering
    \caption{Stellar and planet properties of the \aum\ system used in this study. When taken from the literature, the reference of each parameter value is indicated in the right-hand column$^{\dagger}$. Note that DI reconstructions are performed assuming a stellar inclination of \idi\,=\,80\degr\ (see Section~\ref{sec:section3}). Mid-transit times are given in TBJD - 2\,457\,000 units.}
    \begin{threeparttable}
    \label{tab:stellar_params}    
    \begin{tabular}{cccc}
        \hline
        Quantity & Parameter &  Value & Reference \\
        \hline
        \multicolumn{4}{l}{\textbf{Stellar parameters}} \\
        Distance & D & 9.7248\,$\pm$\,0.0046\,pc & \citetalias{gaia2018}  \\
        Effective temperature   & \teff  & 3700\,$\pm$\,50\,K  & \citetalias{afram2019}  \\
        Surface gravity  & \lgg  & 4.39\,$\pm$\,0.03 & From \rs\ and \mass  \\
        Stellar radius  & \rs  & 0.75\,$\pm$\,0.03\,\rsun & \citetalias{white2019}  \\
        Stellar mass    & \mass & 0.50\,$\pm$\,0.03\,\msun  & \citetalias{plavchan2020} \\
        Luminosity    & \lstar &  0.09\,$\pm$\,0.02\,\lsun & \citetalias{plavchan2009} \\
        Age   & -- & 22\,$\pm$\,3\,Myr  & \citetalias{mamajek2014} \\
        Rotation period   & \pr & 4.8571$^{+0.0037}_{-0.0027}$\,d  & \citetalias{zicher2022} \\
        Projected rotational velocity   & \vs & 7.8\,$\pm$\,0.3\,\kms & \citetalias{klein2021} \\
        \hline
        \multicolumn{4}{l}{\textbf{Planet b}} \\
        Orbital period & \porbb & 8.463000\,$\pm$\,0.000002\,d & \citetalias{martioli2021} \\
        Semi-major axis & \apb & 0.0645\,$\pm$\,0.0013\,au  & \citetalias{martioli2021} \\
        Mid-transit time & \ttrb & 1330.39051\,$\pm$\,0.00015\,d [BJD] & \citetalias{martioli2021} \\
        Planet radius & \rpb & 4.07\,$\pm$\,0.17\,\rearth & \citetalias{martioli2021}  \\
        Velocity semi-amplitude & \kpb & 5.8\,$\pm$\,2.5\,\ms & \citetalias{zicher2022} \\
        Planet mass & \mpb & 11.7\,$\pm$\,5.0\,M$_{\odot}$ & \citetalias{zicher2022}  \\
        Bulk density & \rhopb & 0.97\,$\pm$\,0.43\,g\,cm$^{-3}$ & \citetalias{zicher2022}  \\
        \hline
        \multicolumn{4}{l}{\textbf{Planet c}} \\
        Orbital period & \porbc & 18.859019\,$\pm$\,0.000016\,d & \citetalias{martioli2021} \\
        Semi-major axis & \apc &  0.1101\,$\pm$\,0.0022\,au & \citetalias{martioli2021} \\
        Mid-transit time & \ttrc & 1342.2223\,$\pm$\,0.0005\,d [BJD] & \citetalias{martioli2021} \\
        Planet radius & \rpc &  3.24\,$\pm$\,0.16\,\rearth & \citetalias{martioli2021} \\
        Velocity semi-amplitude & \kpc & 8.5\,$\pm$\,2.5\,\ms & \citetalias{zicher2022} \\
        Planet mass & \mpc & 22.2\,$\pm$\,6.7\,M$_{\odot}$  & \citetalias{zicher2022} \\
        Bulk density & \rhopc & 3.66\,$\pm$\,1.28\,g\,cm$^{-3}$ & \citetalias{zicher2022} \\
        \hline
    \end{tabular}
     \begin{tablenotes}
     \footnotesize
     \item[$\dagger$] To gain some space in the table, we use aliases for the references. \citetalias{gaia2018}, \citetalias{afram2019}, \citetalias{white2019}, \citetalias{plavchan2020}, \citetalias{plavchan2009}, \citetalias{mamajek2014}, \citetalias{zicher2022}, \citetalias{klein2021} and \citetalias{martioli2021} stand respectively for \citet{gaia2018}, \citet{afram2019}, \citet{white2019}, \citet{plavchan2020}, \citet{plavchan2009}, \citet{mamajek2014}, \citet{zicher2022}, \citet{klein2021} and \citet{martioli2021}.
    \end{tablenotes}
    \end{threeparttable}   
\end{table*}

\aum\ was observed between November 2020 and September 2021 (ESO Periods 106 \& 107) with the  high-accuracy radial velocity planet searcher \citep[HARPS;][]{mayor2003}, mounted at the ESO's 3.6-m telescope at La Silla observatory\footnote{Observing programs 105.20N0.001 (PI: Zicher) and 106.21MA.001 (PI: Aigrain)}. A total of 91 high-resolution  spectra ($\mathcal{R}$\,=\,115\,000) were collected in high-accuracy mode: 22 between 2020 November 15 and December 9, and 69 between 2021 May 24 and September 22. The three observations collected on 2021-08-23, which exhibit strong contamination by the Moon (illumination larger than 99\%) along with degraded seeing with full width at half maximum (FWHM) larger than 4\arcsec, are discarded in the following analysis. The remaining observations were taken in good seeing conditions with FWHM of ranging between 0.6\arcsec\ and 2.3\arcsec\ with a median value of 1.1\arcsec\ and with an airmass systematially lower than 2. The full journal of observations is given in Table~\ref{tab:journal_obs}.

\begin{figure}
    \centering
    \includegraphics[width=\linewidth]{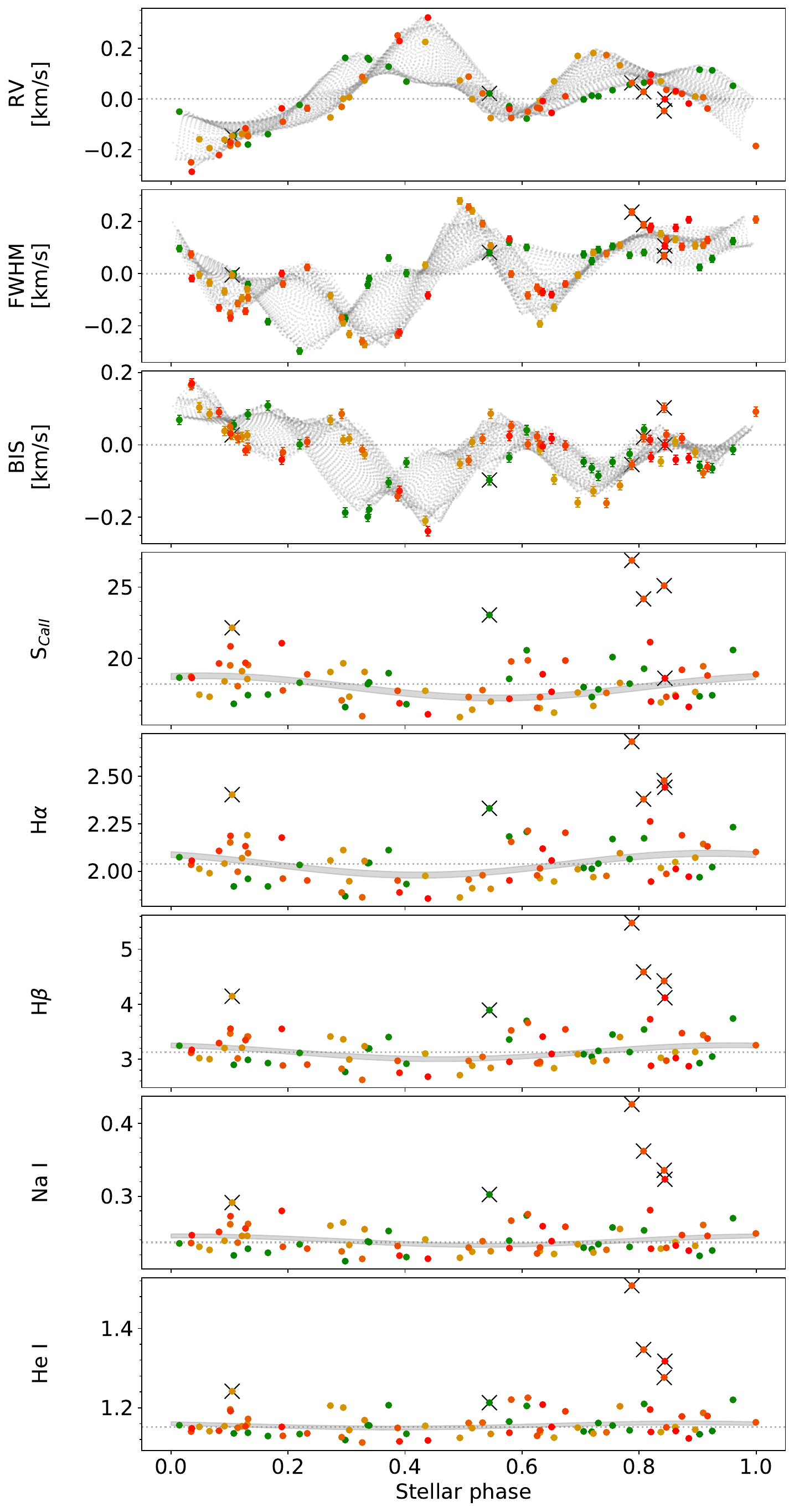}
    \caption{From top to bottom: time series of DRS-provided RV, FWHM, BIS and chromospheric emission indicators based on Ca\,II H \& K, H$\alpha$, H$\beta$, Na\,I and He\,I lines, folded at a rotation period of 4.86\,d. In each panel, data points of different colors belong to the same stellar rotation cycles (dark green and dark red depict the oldest and most recent observations, respectively). The horizontal dotted lines mark the mean of the data. The gray lines indicate (i)~the best-fitting GP prediction in the first three panels (see the description of the fitting procedure in Section~\ref{sec:sec3.2}) and (ii)~the best-fitting sine-wave (using a least-squares estimator at a period of 4.86\,d). Points marked by a black cross were flagged as affected by a stellar flare in Sec.~\ref{sec:section3.1}.}
    \label{fig:indic_all}
\end{figure}

The data were reduced using version 3.8 of HARPS data reduction software (DRS). The reduced spectra feature signal-to-noise ratios (S/N) per pixel at 550\,nm ranging from 45 to 139 with a median value of 102. DRS RVs were derived from densely-sampled cross-correlation functions (CCFs) with a velocity bin of 0.25\,\kms\ and feature a dispersion 121\,\ms\ RMS with median uncertainties of 5.2\,\ms\ RMS. Two indicators of the line shape, namely the FWHM and bisector inverse slope \citep[BIS;][]{queloz2001}, were also provided by the DRS (see the detailed description of the RV extraction in the section~2 of \citetalias{zicher2022}).

For the Doppler imaging inversion process (Sec.~\ref{sec:section4}), we compute CCFs of 0.85\,\kms\ velocity bin (corresponding to the velocity bin of HARPS spectra) for each observation using a M2 line mask optimized for the HARPS spectral range \citep{bonfils2013}. Each CCF was truncated outside of a $\pm$11\,\kms\ window from the line center, and residual slopes in the continuum were corrected using a linear fit to the extreme profile wings. As a sanity check, we manually computed the RVs and FWHMs respectively from the first- and second-order momemts of the normalised profiles. The resulting time series differ by no more than $\sim$5\,\ms\ RMS from their DRS counterparts, confirming that the normalisation process has no more than a marginal impact on the centroid of the CCFs for \aum.

In what follows, we define the stellar rotational phase using the stellar rotation period \pr\,=\,4.86\,d reported in \citet{plavchan2020} and the reference date corresponding to the mid-transit time of \aum\,b (BJD = 2458651.993) also used in the spectropolarimetric study carried out in \citet{klein2021}. All the system parameters used in this study are listed in Table~\ref{tab:stellar_params}.

\section{Activity indicators}\label{sec:section3}

In this section, we investigate the non-radiative chromospheric emission of \aum\ from our time series of HARPS spectra. Optical chromospheric lines are indeed well-known indicators of the surface distribution of the small-scale magnetic field for low-mass stars and correlate generally well with stellar activity RV signals \citep{bonfils2007,boisse2009,gomes2011}. In addition, these lines are the best benchmarks for identifying which of our observations are affected by stellar flares in the absence of simultaneous high-cadence photometric monitoring. Finally, the non-radiative chromospheric emission flux is expected to be one of the best proxies of star-planet interactions \citep[e.g.,][]{saar2001,fares2010,cauley2018,strugarek2019,cauley2019}.

\subsection{Chromospheric emission indices}\label{sec:section3.1}

We compute indices based on the flux in the core of Ca\,II H \& K (resp. 3968.47 \& 3933.66 {\AA}, hereafter S$_{\rm{CaII}}$), H$\alpha$ (6562.808 {\AA}), H$\beta$ (4861.363 {\AA}), Na\,I D1 \& D2 (resp. 5895.92 \& 5889.95 {\AA}) and He\,I D3 (5875.62\,{\AA}, hereafter He\,I) chromospheric lines. As expected from the late spectral type and intense magnetic activity of \aum, all these lines exhibit a strong emission reversal, variable from one night to the other (see the full line profiles in Fig.~\ref{fig:detail_chromospheric_lines}). We used the method of \citet{zechmeister2017} to compute chromospheric emission indices from the blaze-corrected HARPS spectra in the stellar rest frame. In each case, the mean flux within the line core is divided by the average flux between two reference continuum regions on both sides of the line. Except for H$\beta$, we used the integration windows defined in \citet{gomes2011} for late-type stars, but broadened to account for the Doppler-broadening of the line given the star's projected equatorial velocity. For H$\beta$, the flux is measured on a 1.2-{\AA} window, and the continuum is estimated from 5-{\AA} windows centered on 4855 and 4870\,{\AA}.

The resulting time series of chromospheric emission indices are folded to the stellar rotation phase in Fig.~\ref{fig:indic_all}. For each chromospheric index, we note that the same handful of points appear almost systematically significantly higher than the typical emission flux. These outliers are most likely attributable to flares occuring shortly before or during the visit. In order to identify which epochs are affected by a flaring event, we compute a master chromospheric index, $C_{\rm{m}}$ by taking the median-normalized sum of all chromospheric emission indices. We then applied a 3$\sigma$-clipping process to the $C_{\rm{m}}$ time-series and flagged 6 epochs likely affected by a flaring event, which are listed in Table~\ref{tab:journal_obs} and marked as black crosses in  Fig.~\ref{fig:indic_all}.

\subsection{Search for periodicities}\label{sec:sec3.2}

\begin{figure*}
    \centering
    \includegraphics[width=\linewidth]{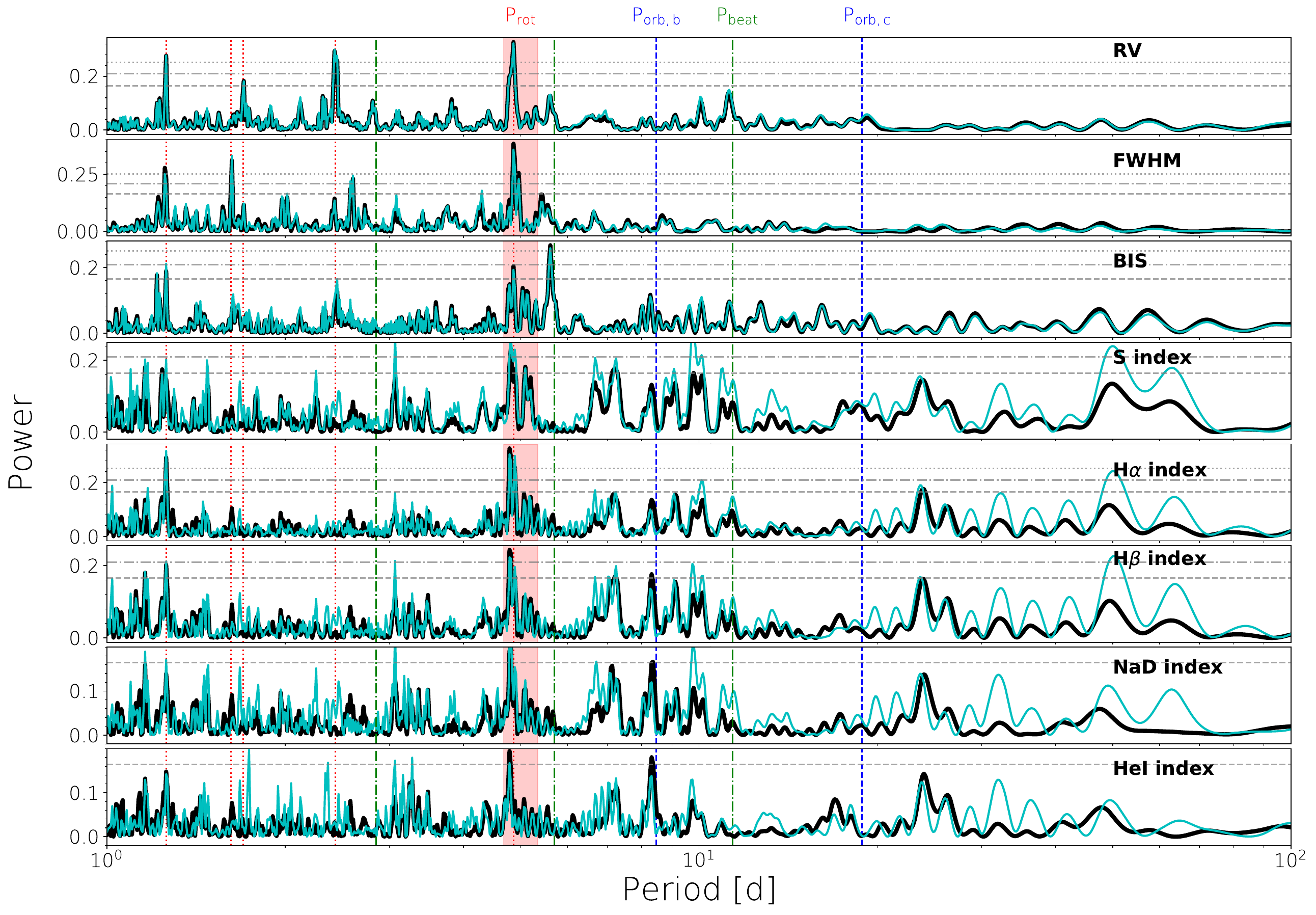}
    \caption{From top to bottom: GLS periodograms of the time series of RVs, FWHMs, BISs, and chromospheric emission indicators based on Ca\,II H \& K, H$\alpha$, H$\beta$, Na\,I and He\,I lines. In each panel, the thick black and thin cyan solid lines respectively show the GLS periodograms of the full data set and of the data set after excluding the observations affected by flares flagged in Section~\ref{sec:section3.1}. The horizontal dashed, dash-dotted and dotted gray lines indicate the false alarm probabilities of respectively 10, 1 and 0.1\%, computed using the method of \citet{baluev2008}. The red vertical dotted lines indicate the stellar rotation period and its harmonics, i.e., from left to right, $1 - f_{\rm{rot}}$, $3 f_{\rm{rot}}$, $1 - 2f_{\rm{rot}}$, $2 f_{\rm{rot}}$ and $f_{\rm{rot}}$, where $f_{\rm{rot}}$\,=\,$1/P_{\rm{rot}}$. The red vertical band delimits the equatorial and polar rotation periods (resp. 4.675\,$\pm$\,0.006\,d and 5.34\,$\pm$\,0.05\,d) measured for the large-scale magnetic field of \aum\ in \citet{klein2021}. The orbital periods of planets b and c are marked by the blue vertical dashed lines. Finally, the green vertical dash-dotted lines indicate the
    the beat period (P$_{\rm{beat}}$) between \aum\,b and its host star as well as its second and fourth harmonics (see Sec.~\ref{sec5:SPMI} and Fig.~\ref{fig:planet signal}). Note that the spectral window of the observations, shown in the Fig.~3 of \citetalias{zicher2022}, does not exhibit any significant peak at the periods of interest. Note also that the periodograms were computed using the \texttt{astropy} python module \citep{astropy2013,astropy2018}.}
    \label{fig:1D_periodogram}
\end{figure*}

As shown in Fig.~\ref{fig:indic_all}, the chromospheric emission flux exhibits clear temporal variations, consistent from one index to the other, throughout the timespan of our observations (see also the line profiles in Fig.~\ref{fig:detail_chromospheric_lines}). However, the chromospheric indices exhibit a significantly weaker modulation with \pr\ and seem to evolve on much shorter timescales than the indicators based on the position and shape of the CCFs (e.g., RV, FWHM or BIS). Such discrepancy is further evidenced by the low correlation observed between each chromospheric index and the RV time-series (see the Pearson correlation coefficients in Table~\ref{tab:modul_indic} and the correlation plot in Fig.~\ref{fig:correl_plot}). This suggests that, in contrast to CCF-based indicators whose variations are primarily dominated by brightness inhomogeneities at the stellar surface, the fluctuations in the non-radiative chromospheric flux are most likely caused by complex combination of phenomena (e.g., active regions at the stellar surface, nanoflares, interactions with external bodies) of similar intensity.

The Generalized Lomb-Scargle periodograms \citep[GLS;][]{zechmeister2009} of the RV, FWHM, BIS, S$_{\rm{CaII}}$, H$\alpha$, H$\beta$, Na\,I, and He\,I time-series are shown in Fig.~\ref{fig:1D_periodogram}. Unsurprinsingly, the periodograms of the CCF-based activity indicators exhibit prominent peaks at \pr\ and its first harmonic. By modelling each of these time series independently using a Gaussian Process \citep[GP;][]{Rajpaul2015} with the quasi-periodic covariance kernel introduced in \citet{Haywood2014} as implemented in the \texttt{pyaneti} python package \citep{barragan2021}, we find consistent rotation periods of 4.863\,$\pm$\,0.004\,d, 4.86\,$\pm$\,0.01\,d and 4.867\,$\pm$\,0.004\,d for the RV, FWHM and BIS time-series, respectively\footnote{Note that the exploration of the parameter space was performed using the method described in Section~3.4 of \citetalias{zicher2022}, i.e. by through a Markov chain Monte Carlo sampler in the Bayesian framework.}. As illustrated in Fig.~\ref{fig:1D_periodogram}, we note that including the flares in the GP fit no more than marginally affects the best-fitting \pr.

\begin{table*}
    \centering
    \caption{Main results of the analysis \aum's chromospheric emission. In lines~1 and~2, we give the best-fitting stellar rotation period for the chromospheric indices and the Pearson correlation coefficient with the RV time-series. In lines~3, we list the best-fitting period at $\sim$8.3\,d (i.e., near the orbital period of \aum\,b) with the associated FAP level. Finally, we give the emitted power associated to this period with and without including the flare-flagged observations (see Sec.~\ref{sec5:SPMI}) in lines~4 and~5.}
    \label{tab:modul_indic}
    \begin{tabular}{cccccc}
        \hline
        \textbf{Chromospheric line} & Ca II  &  H$\alpha$ & H$\beta$ & Na\,I & He\,I  \\
        \hline
        \pr\ & 4.81\,$\pm$0.01\,d & 4.80\,$\pm$0.01\,d & 4.80\,$\pm$0.01\,d & 4.81\,$\pm$0.01\,d & 4.79\,$\pm$0.01\,d \\
        Correlation coef. w/ RV &  -0.336 & -0.245 & -0.258 & -0.297 & -0.145 \\
        \porbb & -- & -- & -- & 8.35\,$\pm$0.04\,d (1\%) &  8.33\,$\pm$0.04\,d (0.2\%) \\
        Power (w/ flares) & -- & -- &  -- & 9.7\,$\pm 2.2 \times 10^{17}$\,W & 5.4\,$\pm 1.3 \times 10^{17}$\,W \\
        Power (w/o flares) & -- & -- & -- & 5.7\,$\pm 1.7 \times 10^{17}$\,W & 2.9\,$\pm 0.9 \times 10^{17}$\,W  \\
        \hline
    \end{tabular}
\end{table*}

The periodogram of each of the chromospheric emission indices exhibits a prominent peak at $\sim$4.8\,d, slightly lower than the photometric stellar rotation period (see Table~\ref{tab:modul_indic}). This suggests that all seven chromospheric emission lines probe similar regions of the stellar atmosphere, and that these regions differ from the main brightness inhomogeneities inducing most of the photometric and CCF variations. We also note that these values lie in between the stellar equatorial and polar rotation periods, respectively equal 4.675\,$\pm$\,0.006\,d and 5.34\,$\pm$\,0.05\,d, measured from the differentially-rotating large-scale magnetic field at the surface of \aum\ in \citet{klein2021}. Each chromospheric emission indicator also exhibits a more-or-less prominent peak at a period of 8.33\,$\pm$\,0.04\,d, $\sim$3$\sigma$ lower than the orbital period of \aum\,b (8.463000\,$\pm$\,0.000002\,d). We also note that, for each chromospheric indicator, including the 6 observations affected by a stellar flare seems to enhance the observed modulation at 8.33\,d. However, since (i)~the significance of this peak reaches no more than a false alarm probability (FAP) of 10\% (for He\,I), and (ii)~other peaks of unclear origin but similar significance are detected at periods of $\sim$7\,d and $\sim$10\,d, we caution than this detection remains marginal and that no robust conclusion can be drawn, at this stage, from the chromospheric emission indices alone.



\subsection{Bidimensional periodogram analysis}\label{sec:sec3.3}

\begin{figure}
    \centering
    \includegraphics[width=\linewidth]{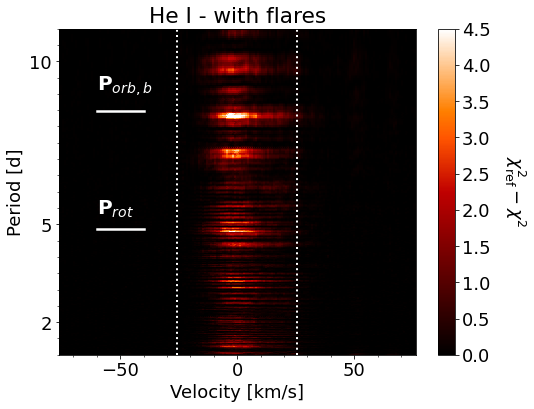}\\
    \includegraphics[width=\linewidth]{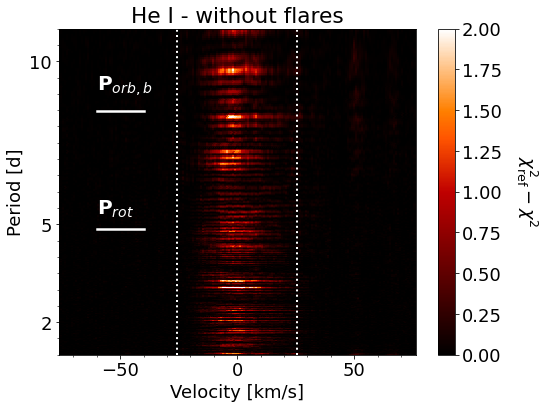}
    \caption{2D periodograms for the He\,I D3 line (5875.62 {\AA}). Each vertical line is a GLS periodogram computed with the \texttt{astropy} python module \citep{astropy2013,astropy2018}. For each column, the color code depicts the difference in reduced-$\chi^{2}$ of the residuals between the best-fitting constant reference model, $\chi_{\rm{ref}}^{2}$, and the best-fitting simple sine-wave model for each period on the Y-axis, $\chi^{2}$. The two white vertical dotted lines mark the blue- and red-hand sides of the line core.}
    \label{fig:2D_period_he}
\end{figure}

In order to further investigate the putative modulation at 8.33\,d unveiled from the chromospheric indices, we compute the bidimensional periodogram of each emission line to quantify the evolution of the flux variability across the line profile, as it is commonly done for young PMS stars \citep[e.g.,][]{johns1995,bouvier2007,sousa2021,finociety2021}. We first normalize the continuum of each chromospheric line from the blaze-subtracted HARPS spectra in the stellar rest frame. For each spectrum, we mask the chromospheric emission line as well as all absorption features with relative depth larger than 5\% in a 3-nm wide window centered on the chromospheric line. Within the window, we iteratively smooth the spectrum using a Savitzky-Golay filter, reject points deviating by more than 3$\sigma$ from the mean of the residuals, and repeat the process until no more outlier is flagged. Finally, we interpolate the resulting smoothed spectrum using a cubic spline kernel, and divide the full spectrum within the window by the interpolated function. As a final step, we divide each continuum-normalised spectrum by the median out-of-flare spectrum and compute a GLS periodogram for each velocity bin of the resulting time series of line profiles using the \texttt{astropy} package \citep{astropy2013,astropy2018}.

The resulting bidimensional periodograms are shown in Fig.~\ref{fig:2D_period_he}, for He I, and in Fig.~\ref{fig:detail_2D_periodograms_fl} (resp. Fig.~\ref{fig:detail_2D_periodograms_nofl}) for the other lines with (resp. without) including the observations affected by stellar flares in the input data. We find that the bidimensional periodogram of the He\,I line exhibits a clear significant peak at a period of 8.33\,$\pm$0.04\,d, well-centered on the line profile, present whether we include the flares in the input data set or not. To estimate the significance of the peak, we integrate each normalized line profile over a 16-\kms\ wide window centered on the emission line, i.e., where the modulation is strongest, and compute the GLS periodogram of the resulting time series. Using a bootstrap-based method to estimate the FAP levels, we find that the peak at 8.33\,d reaches FAP levels of 0.2\% and 1\% when the flares are respectively included and excluded from the data set. As a sanity check, we built 100 data sets by removing a third of the data points randomly chosen within the observational dates and found that the peak was systematically detected at a confidence larger than 2$\sigma$, reinforcing the non-spurious nature of this detection. As a final test, we created two subsets C$_{1}$ and C$_{2}$ of data from the time series of continuum-normalised He I profiles, containing the data collected at planet orbital phases\footnote{The orbital phase of \aum\,b is defined using the planet mid-transit time given in Sec.~\ref{sec:section2} (i.e., BJD = 2458651.993) and the planet orbital period given in Table~\ref{tab:stellar_params} (i.e., 8.463\,d).} respectively between 0.75 and 0.25 (when the planet is closer from the observer than the star) and between 0.25 and 0.75 (i.e., when the star is closer from the observer than the planet). The bidimensional periodograms of C$_{1}$ and C$_{2}$ are shown in Fig.~\ref{fig:detail_2d_orb_phase}. We find that the peak at 8.33\,d almost completely disappears in case C$_{2}$, whereas the signal is still clearly identified (even though noisier) in case C$_{1}$, which corresponds to what one would have expected if the observed modulation is indeed due to a planet-induced hot spot in the stellar atmosphere. We however caution that, since the selected subsets C$_{1}$ and C$_{2}$ are based on the planet orbital period, the search for peaks around a 8.33-d period in each of these subsets incurs the risk of aliasing, and that no firm conclusion can be drawn from this analysis alone.

As shown in Fig.~\ref{fig:detail_2D_periodograms_fl} and~\ref{fig:detail_2D_periodograms_nofl}, a prominent peak at 8.33\,d is similarly identified from the Na I lines but the latter is only tentatively detected when the observations affected by stellar flares are excluded from the data set. This modulation remains no more than marginally detected from the H$\alpha$ and H$\beta$ line profiles, whose bidimensional periodograms are dominated by the stellar rotational modulation. Similarly, no detection can be claimed from the Ca H \& K lines which, as expected from Fig.~\ref{fig:1D_periodogram}, exhibit chaotic bidimensional periodograms with multiple peaks between 7 and 10\,d of unclear origin. While the late spectral type of AU\,Mic means that the continuum SNR is relatively low at the around the Ca H \& K lines, it is not clear that this alone explains the apparently noiser periodogram for those lines.  The continuum-normalised emission flux within the Ca H \& K lines does not exhibit any correlation with the airmass or the SNR of the observations, which would have been expected if the lines were affected by atmopsheric dispersion or low-signal noise.


\section{Brightness Imaging}\label{sec:section4}

\subsection{Description of the method}\label{subsec:4.1}

In this section, we propose to jointly recover both the stellar brightness topologies and the planet orbital parameters from the time series of HARPS CCFs using Doppler imaging \citep[e.g.,][]{vogt1987,semel1989,brown1991,donati1997b,donati2000,donati2014}. Our DI code models the stellar surface as a dense grid of 5\,000 cells, each featuring a relative brightness coefficient (resp. 1, $>$1 and $<$1 for the quiet photosphere, bright plage and dark spot). At each observation date, local intensity line profiles are computed for each cell in the visible hemisphere of the star using Unno-Rachkovsky's analytical solution to the radiative transfer equations in a Milne-Eddington atmosphere \citep{unno1956}. These local line profiles are then (i)~Doppler-shifted according to the local RV of the cell, (ii)~weighted by the local limb-darkening, sky-projected area and local brightness, and (iii)~combined into global line profiles and compared to the observed CCFs. In practice, we use a conjugate gradient algorithm to iteratively compute brightness togologies until the time series of synthetic line profiles matches the observed CCFs down to a user-provided \chisqr. The degeneracy of the inversion problem is overcome by imposing a maximum-entropy regularisation condition following \citet{skilling1984}, i.e., by choosing the brightness distribution with the minimum spot coverage.

In moderately-fast rotators like \aum, the unknown intrinsic profile of the star still significantly contributes to the width and depth of the observed line profiles. In order to ensure that the DI code focuses on reconstructing activity-induced profile distortions rather than systematic differences between synthetic and observed line profiles \citep[see][]{hebrard2016}, we use the iterative procedure described in \citet{klein2021}. We first use our DI code to perform a maximum-entropy inversion of the observed line profiles $\boldsymbol{I_{\rm{obs}}}$, which provides a time series of synthetic profiles $\boldsymbol{I_{\rm{syn}}}$. We then subtract the median difference between $\boldsymbol{I_{\rm{obs}}}$ and $\boldsymbol{I_{\rm{syn}}}$ from $\boldsymbol{I_{\rm{obs}}}$ and repeat the process until the median difference between $\boldsymbol{I_{\rm{obs}}}$ and $\boldsymbol{I_{\rm{syn}}}$ is flat (which takes typically 5-10 iterations). Similarly to \citet{klein2021}, we find that this procedure marginally affects the profile RVs for \aum. This is due to the fact that both stellar rotation and close-in planet orbital periods are significantly lower than the total time span of the data. As a consequence, both profile distortions (due to stellar activity) and shifts (mostly due to the close-in planets) are averaged out in the median difference between $\boldsymbol{I_{\rm{obs}}}$ and $\boldsymbol{I_{\rm{syn}}}$.

We then adapt the method introduced in \citet{petit2015} and applied in e.g., \citet{donati2017,yu2017,klein2021,heitzmann2021}. We simultaneously search for planet-induced shifts in \aum’s CCF time-series whilst correcting for their activity-induced distortions using DI. Consistently with the transit analyses, we model each planetary RV signal, $V_{\rm{p}}$, assuming that the planet orbits are circular, such that

\begin{eqnarray}
V_{\rm{p}} (t) = K_{\rm{p}} \sin \left[ 2 \pi\:  \frac{T_{0} - t}{P_{\rm{orb}}} \right],
\label{eq:rv_p}
\end{eqnarray}

\noindent
where $K_{\rm{p}}$ is the semi-amplitude of the planet RV signature, and $T_{0}$ and $P_{\rm{orb}}$ are the planet mid-transit time and orbital period. In the fiducial case, we assume that \aum\,b and \aum\,c are perfectly phased with their respective transit curves, leaving their velocimetric semi-amplitudes as the only free planet parameters. Our DI code includes a solar-like model of differential rotation (DR) as described in \citet{donati2000} and \citet{petit2002}. The evolution of stellar rotation rate $\Omega$ as a function of the colatitude $\theta$ is described by

\begin{eqnarray}
\Omega (\theta) = \Omega_{\rm{eq}} - \rm{d}\Omega \cos^{2} \theta,
\label{eq:dr}
\end{eqnarray}

\noindent
where \oeq\ is the rotation rate at the stellar equator and \dome\ is the difference of rotation rate between the stellar equator and the pole. We carry out the inversion of the observed CCFs for a wide range of planet and DR parameters. For each set of parameters, the line profiles are shifted according to the planet RV signature given by Eq.~\ref{eq:rv_p} and then reconstructed with DI up to a given level of entropy (i.e., spot coverage), thus yielding a value of \chisqr\ quantifying the goodness of DI fit. The best parameters and their error bars are estimated afterwards from the \chisqr\ map using \citet{press1992}'s statistics.

\subsection{Application to AU Mic}\label{sec:sec4.2}

\begin{figure}
    \centering
    \includegraphics[width=\linewidth]{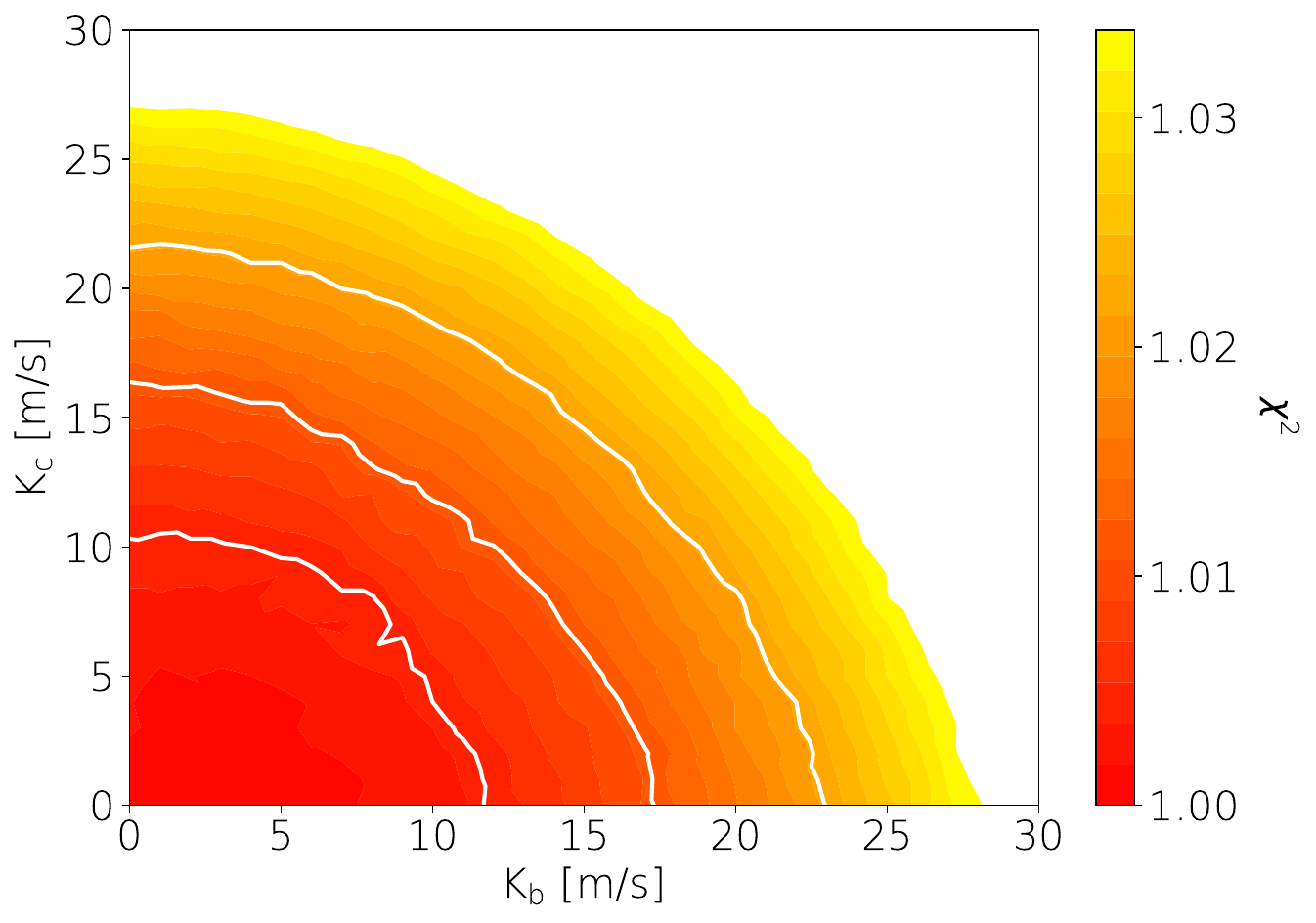} \\
    \includegraphics[width=\linewidth]{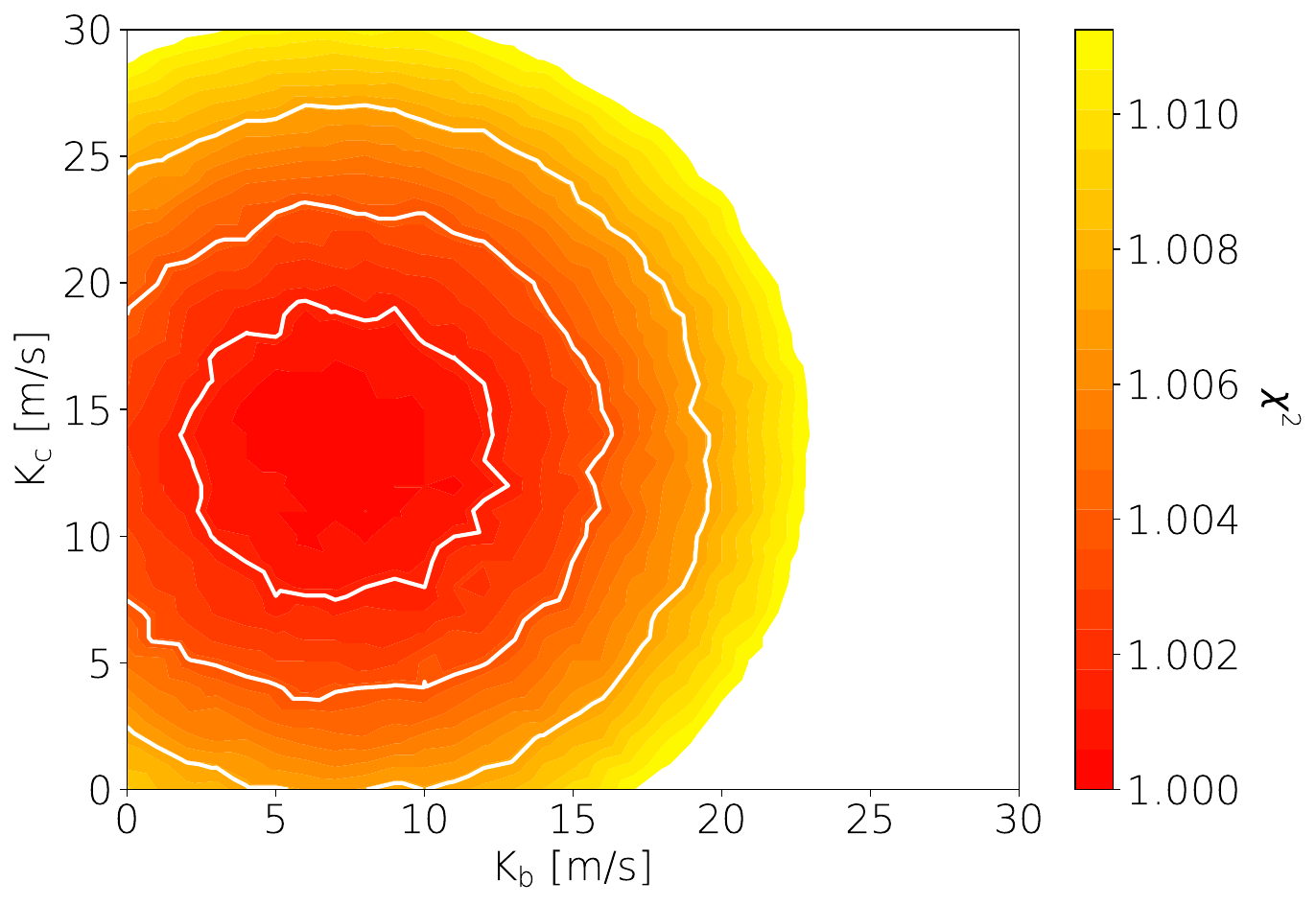} \\    
    \caption{2D \chisqrr\ maps obtained by performing brightness reconstructions of the time series of HARPS CCFs collected over the November-December 2020 period (top panel) and over the May-September 2021 period (bottom panel). The DI reconstructions shown here are performed assuming the best-fitting DR parameters obtained for each period (see Sec.~\ref{sec:sec4.2}). In each panel, the 1, 2 and 3$\sigma$ levels are depicted by white solid lines. }
    \label{fig:chi2_maps}
\end{figure}

Intrinsic stellar variability is not yet included in the current version of our DI code. We thus divide our time series of HARPS CCFs into two subsets of data, $\boldsymbol{I_{2020}}$ and $\boldsymbol{I_{2021}}$, containing respectively the 21 observations taken in November/December 2020 and the remaining 61 observations collected in 2021, after removing Moon- and flare-contaminated line profiles. We then perform the DI analysis independently on each dataset. These two subsets of data respectively span 24 and 121\,d (i.e., resp. 5 and 25 stellar rotation cycles), lower or of similar length to the $\sim$100-d time scale on which \aum's activity signal evolves \citep{plavchan2020}. In each DI reconstruction, we adopt a linear limb-darkening law of coefficient 0.66 \citep{claret2011}, an axial inclination of \idi\,=\,80\degr\footnote{Note that, although the axial inclination of \aum\ is likely close to 90\degr, imposing a stellar inclination of 80\degr\ in the DI reconstruction allows one to prevent any north-south degeneracy. Note that, as shown for example in \citet{petit2015}, variations of stellar inclination $\la$\,10\degr\ marginally affect the recovered brightness topology and planet orbital parameters.}, and a \vs\ of 7.8\,\kms\ \citep[see][]{weise2010,klein2021}.

\begin{figure}
    \centering
    \includegraphics[width=\linewidth]{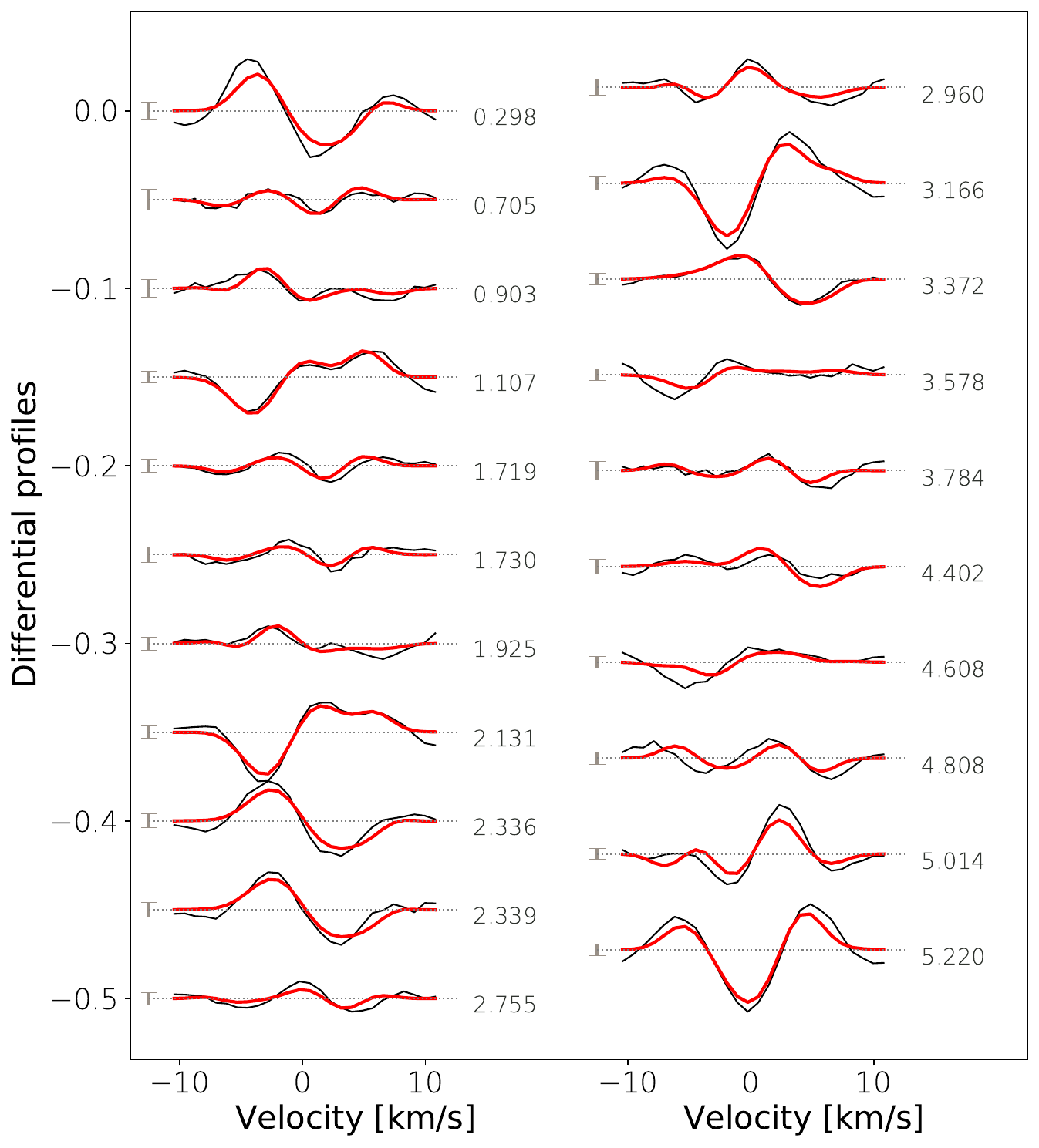}
    \caption{Time series of observed (black thin lines) and synthetic (red thick lines) mean-subtracted line profiles (a.k.a. differential profiles) collected in 2020. The stellar rotation cycle is written at the right-hand side of each profile whereas the $\pm$\,1\,$\sigma$ uncertainties on the line continuum is indicated at the left-hand side of the profiles. Note that the stellar rotational cycles are counted from the cycle corresponding to the first observation collected over the 2020 period.}
    \label{fig:prof_fit_aumic_2020}
\end{figure}

We apply the process described in Section~\ref{subsec:4.1} to independently invert $\boldsymbol{I_{2020}}$ and $\boldsymbol{I_{2021}}$ into surface brightness topologies and planet velocimetric semi-amplitudes. We find that the 2020 data set is consistent with a solid-body rotation at a rotation period of 4.85\,$\pm$\,0.01\,d, in line with the photometric rotation period 4.86\,d (see Table~\ref{tab:stellar_params}). In a similar way, the 2021 profiles are found to be sheared by a weak DR with \oeq\,=\,1.294\,$\pm$0.001\,\radd\ and \dome\,=\,0.013\,$\pm$\,0.005\,\radd, implying rotation periods of 4.856\,$\pm$\,0.004\,d and 4.90\,$\pm$\,0.03\,d at the equator and pole, respectively.

The reduced \chisqr\ maps are shown in the ($K_{\rm{b}}$, $K_{\rm{c}}$) space at the best-fitting DR parameters for both data sets in Fig.~\ref{fig:chi2_maps}. Unsurprisingly, the relatively low number of data points on the 2020 period (21 profiles of 26 data points each) makes it impossible to recover any of the two planets from the $\boldsymbol{I_{2020}}$ data set alone. In contrast, we obtain a 3.3-$\sigma$ detection of a planetary signal of
semi-amplitude of $K_{\rm{c}}$\,=\,13.3\,$\pm$\,4.1\,\ms\ at the orbital period of \aum\,c, whereas the RV signature of \aum\,b remains no more than marginally recovered, with a semi-amplitude of $K_{\rm{b}}$\,=\,7.1\,$\pm$\,3.8\,\ms. To ensure that the recovered signals are indeed of planetary origin, we reapplied the DI process to the 2021 data set but, this time, freezing the DR parameters to their best estimates and leaving the two orbital periods and velocimetric semi-amplitudes of the planets as free parameters of the model. We recovered orbital periods of 8.47\,$\pm$\,0.01\,d and 18.87\,$\pm$\,0.02\,d for \aum\,b and \aum\,c, both fully consistent with the literature values (see Table~\ref{tab:stellar_params}). Moreover, the planet semi-amplitude estimates remain similar, but slightly less precise than their counterparts from the previous paragraph.

\begin{figure}
    \centering
    \includegraphics[width=\linewidth]{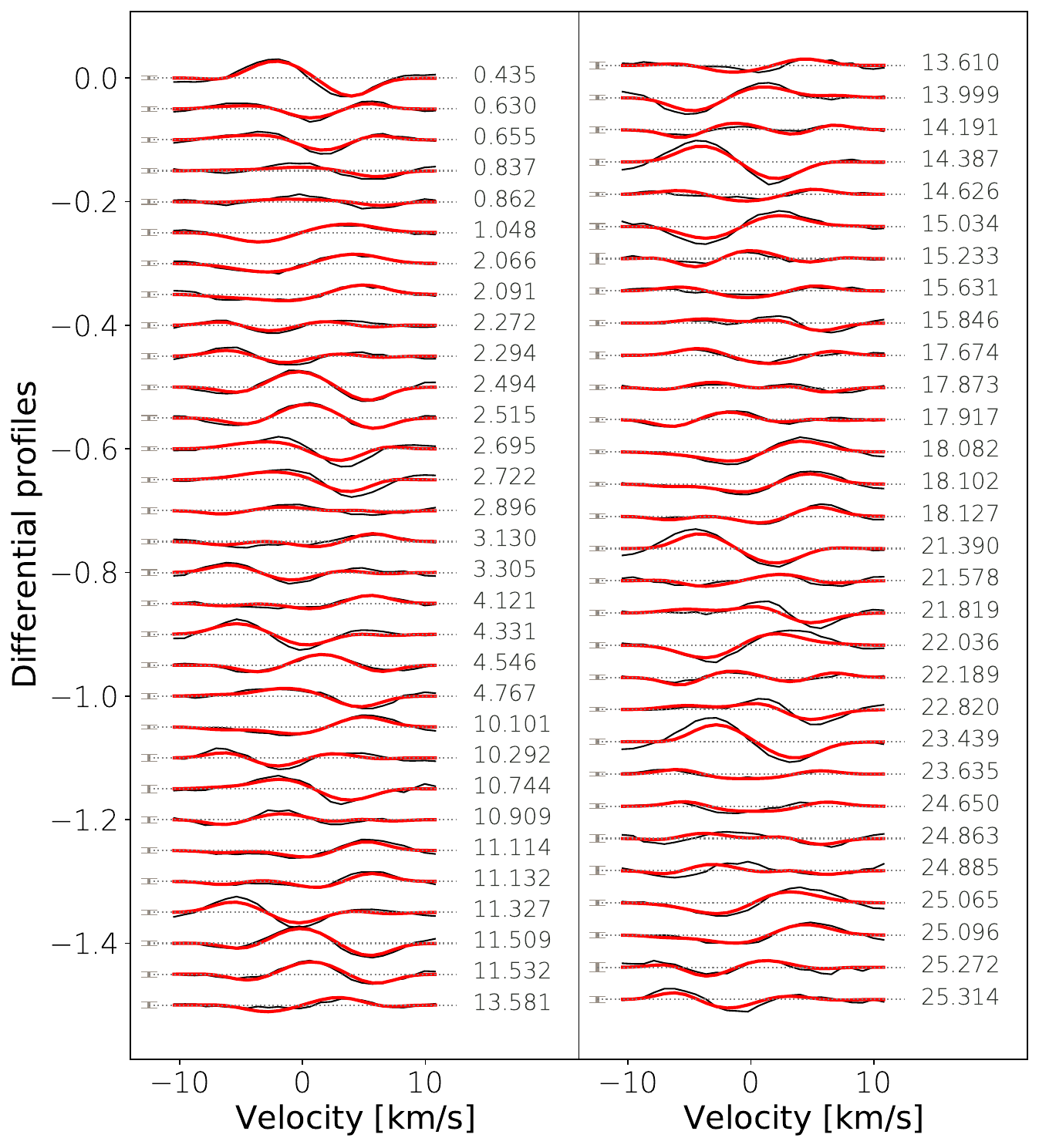}
    \caption{Same as Fig.~ \ref{fig:prof_fit_aumic_2020}, but, this time, for the observations collected on the 2021 period. Note that the stellar rotational cycles are counted from the cycle corresponding to the first observation collected over the 2021 period.}
    \label{fig:prof_fit_aumic_2021}
\end{figure}
\begin{figure}
    \centering
    \includegraphics[width=\linewidth]{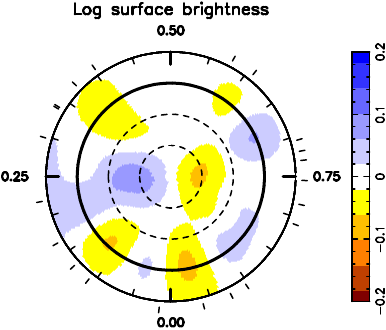}\\
    \includegraphics[width=\linewidth]{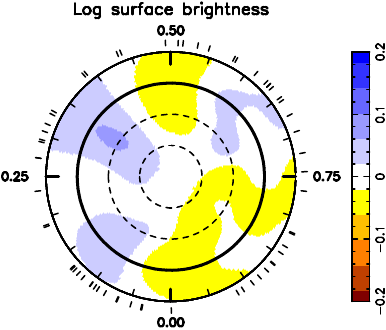}
    \caption{Best-fitting brightness topologies of \aum\ from 2020 (top panel) and 2021 (bottom panel) data sets. Each panel is a flattened polar view of the surface of the star where the color scale depicts the logarithm of the relative surface brightness. The stellar pole is located at the center of the plot whereas the concentric circles indicate the stellar equator (solid line) and $-$30\degr, 30\degr\ and 60\degr\ parallels. The radial ticks around the star mark the phases of observations.}
    \label{fig:maps_aumic}
\end{figure}

The maximum-entropy fits to the HARPS CCFs, shown in Fig.~\ref{fig:prof_fit_aumic_2020} and Fig.~\ref{fig:prof_fit_aumic_2021} for $\boldsymbol{I_{2020}}$ and $\boldsymbol{I_{2021}}$, respectively, reproduce the observations down to \chisqrr\,=\,1. The corresponding brightness topologies, respectively shown in the top and bottom panel of Fig.~\ref{fig:maps_aumic}, feature consistent spot coverages of 4.7 and 5.1\% (typical error bar of 0.5\%). Both maps exhibit more dark spots than bright plages, all preferentially located near the stellar equator. We note that the reconstructed features appear more spread out in 2021 than 2020, which is most likely attributable to the intrinsic evolution of the magnetic activity throughout the 2021 observations.

We now apply the process described in Section~\ref{subsec:4.1} to the whole time series of HARPS CCFs, from November 2020 to September 2021 (82 observations). By doing so, our DI fit relies on more observational constraints and, thus, might yield more precise mass estimates for the two planets, provided that the brightness distribution has not radically changed throughout the observations. We impose a solid-body rotation in the DI reconstruction, to keep consistency with the results individually obtained on both data sets. Unsurprinsingly, our best DI reconstruction cannot reproduce the observed profile to a reduced \chisqr\ lower than $\sim$2, which is most likely attributable to the evolution of stellar activity over the last year. We barely recover the velocimetric semi-amplitudes of both planets, confirming that the intrinsic evolution of the stellar activity is the limiting factor on the accuracy of our planet mass estimates.

\section{Chromospheric activity and star-planet interaction}\label{sec:section5}

In this study, we conducted a spectroscopic analysis of the magnetic activity of \aum\ from an intensive 10-month monitoring campaign with HARPS. In this section, we compare our results to the recent literature and discuss their implications on stellar activity, planet formation and evolution and star-planet interactions.

\subsection{Chromospheric activity and star-planet interactions}\label{sec5:SPMI}

CCF-based activity indicators (i.e., RV, FWHM, BIS) show a clear rotational modulation, slowly-evolving from November 2020 to September 2021, which echoes the brightness distributions obtained with DI (see Fig.~\ref{fig:maps_aumic}). As expected from the young age and late spectral type of \aum, each of the Ca\,II H \& K, H$\alpha$, H$\beta$, Na\,I D1 \& D2 and He\,I D3 chromospheric lines presents a strong time-evolving emission reversal. The non-radiative emission flux in each line appears modulated at a typical period of 4.80\,d, 6-$\sigma$ shorter than the photometric/velocimeter stellar rotation period (4.86\,d). This discrepancy is discussed in Sec.~\ref{sec:section5_DR}. We also note that the non-radiative chromospheric flux also exhibits significantly more short-term variability than the CCF-based indicators, suggesting that the observed emission is resulting from a combination of different physical processes.

\subsubsection{Origin of the 8.33-d moduation}

\begin{figure}
\centering
\includegraphics[width = \columnwidth]{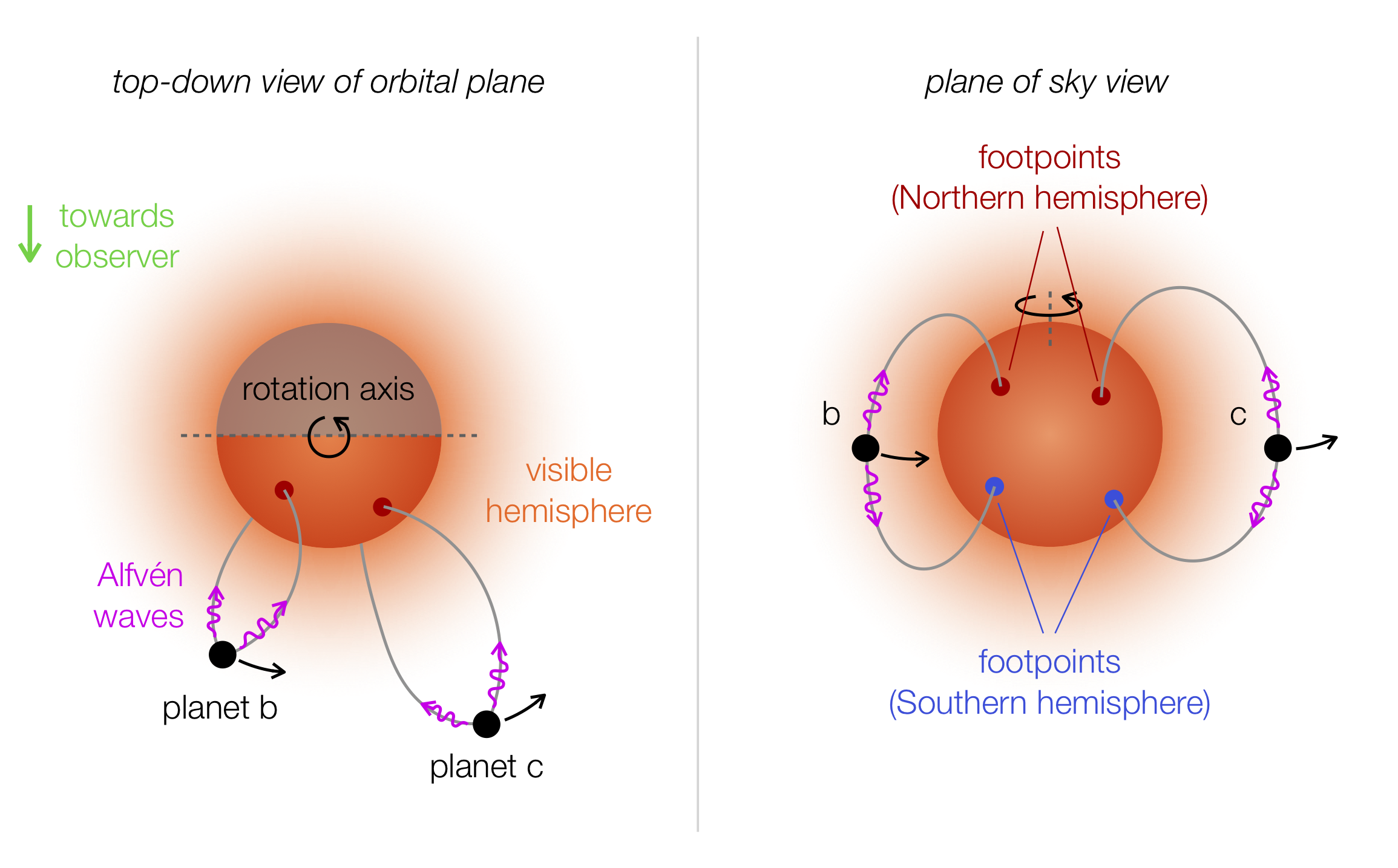}
\caption{Sketch of the scenario where regions of enhanced emission are induced on the stellar surface of AU Mic via a sub-Alfv\'enic interaction between the star's magnetic field and the orbiting planets. In a sub-Alfv\'enic orbit, a planet can produce Alfv\'en waves that carry energy back towards the stellar surface along the field line connecting to the planet. If the point where the field line connects to the stellar surface (footpoint) appears on the hemisphere visible to the observer (left), enhanced chromospheric emission may appear. As planets b and c both orbit in the closed-field region of the star's magnetic field \citep[see][]{kavanagh2021}, they may each produce up to two visible footpoints on the stellar disk (right).}
\label{fig:sketch planet}
\end{figure}

Each chromospheric emission index exhibits a more-or-less prominent modulation at a period of 8.33\,$\pm$\,0.04\,d, most robustly detected from the bidimensional periodogram of the He\,I line (FAP of 0.2\%). This emission signal could potentially be induced by star-planet magnetic interactions (SPMI) between planets b and/or c and its host star. From the large-scale magnetic topology of the star published in \citet{klein2021}, \citet{kavanagh2021} have shown that, for a relatively low mass-loss rate of 27 $\dot{M}_{\sun}$ \citep[][]{plavchan2009}, planets b and c orbit in the sub-Alfv\'enic region of the stellar wind\footnote{Note that he solar mass loss rate $\dot{M}_{\odot}$ is assumed equal to 2\,$\times$\,$10^{-14}$\,M$_{\odot}$\,yr$^{-1}$ \citep{wood2005}.}. In this configuration, the interaction between the planet and the stellar magnetic field is expected to produce Alfv\'en waves, which travel from the planet to the star along the field line connecting the two bodies, inducing potentially-observable signatures in the stellar atmosphere \citep[e.g.,][]{cuntz2000,ip2004,strugarek2019,fischer2019}. A sketch of this scenario is shown in Figure~\ref{fig:sketch planet}.

The power $P$ of the Alfv\'en waves generated in this interaction is \citep{saur2013}
\begin{eqnarray}
P = \pi^{1/2} {R_\textrm{p}}^2 B \rho^{1/2} \Delta u^2 \sin^2 \theta,
\label{eq:planet-induced power}
\end{eqnarray}
where $R_\textrm{p}$ is the planet radius, and $B$ and $\rho$ are the stellar wind magnetic field strength and density at the position of the planet respectively. $\Delta u$ is the relative velocity between the stellar wind and planetary orbit, and $\theta$ is the angle between the vectors $\boldsymbol{B}$ and $\Delta\boldsymbol{u}$. If the observed modulation is induced by the deposition of this energy at the footpoint of the field lines connecting one or both planets to the star, one would expect the observed signal to be modulated at the orbital period of the planet \citep{cuntz2000,saar2001,strugarek2018,fischer2019}. This theoretical explanation is strengthened by the fact that the observed modulation disappears when the orbital phases of planet b lie between 0.25 and 0.75. However, the systematic 0.13-d difference between the observed modulation period and \porbb\ is hard to reconcile with standard SPMI models. As outlined by \citet[][hereafter \citetalias{fischer2019}]{fischer2019}, in the ideal case of a stellar magnetic dipole nearly aligned with the stellar rotation axis and perpendicular to the planet orbital plane, star-planet interactions are likely to induce observable signatures modulated at the beat period of the system, P$_{\rm{beat}}$, given by

\begin{eqnarray}
P_{\rm{beat}} = \frac{2 \pi}{\Omega_{\rm{star}} - \Omega_{\rm{p}}},
\label{eq:pbeat}
\end{eqnarray}

\noindent
where $\Omega_{\rm{star}}$ and $\Omega_{\rm{p}}$ are the respective angular velocities of the star and planet.

This effect can be investigated further using the 3D magnetohydrodynamical simulations of \aum's stellar wind presented in \citet{kavanagh2021}, based on the magnetic topology obtained from the 2019 SPIRou observations of the star \citep{klein2021}. We use the stellar wind model of AU Mic which predicts that both planets b and c are predominantly in sub-Alfv\'enic orbits, corresponding to a mass-loss rate of $27~\dot{M}_{\odot}$. With this model, we compute the positions of the footpoints of the stellar magnetic field lines connecting from planets b and c to the stellar surface, for 500 stellar rotations beginning at the reference time for cycle phase zero of BJD = 2458651.993 \citep{klein2021}. This provides us with both full temporal coverage of our HARPS observations, and also allows us to assess the long-term evolution of the simulated planet-induced signal. At this reference time, the orbital phases of planets b and c are 0.91 and 0.32 respectively.

\begin{figure}
\centering
\includegraphics[width = \columnwidth]{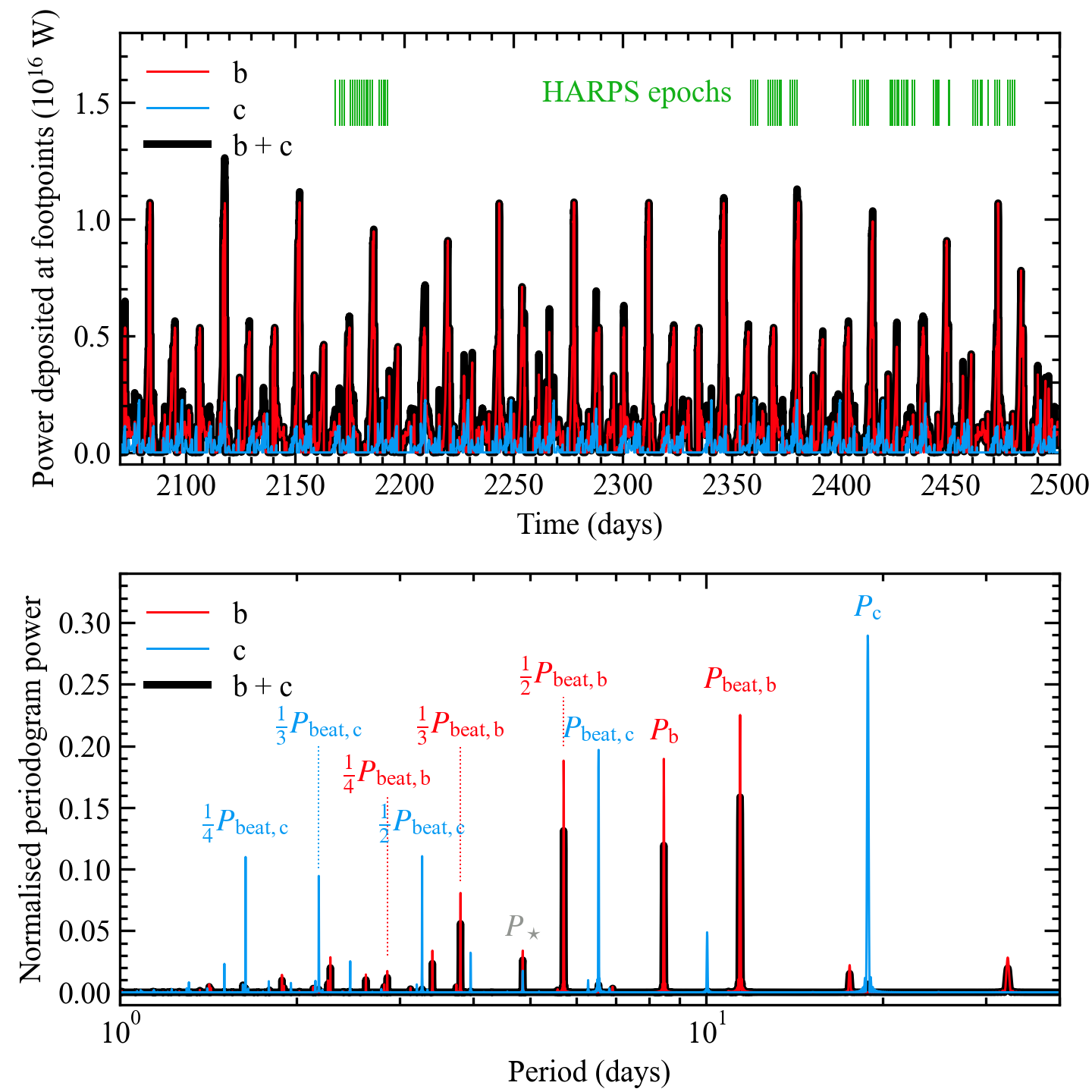}
\caption{\textit{Top:} Power deposited at the footpoints of the stellar magnetic field lines connecting to planets b and c during the HARPS observations, computed using the simulated stellar wind environment of AU Mic presented in \citet{kavanagh2021}. The times are given in TBJD $-\,2\,457\,000$. The thin red/blue lines show the individual contribution of planets b and c respectively, and the thick black line shows the sum of the two signals. Vertical green lines illustrate the epochs of the HARPS observations. \textit{Bottom}: GLS periodograms computed for the simulated planet-induced signal. Again, we show the results for the individual (thin red/blue lines) and sum of the two signals (thick black line). The signals induced by planets b and c show strong periodicity at their respective orbital periods, as well as the fundamental and harmonics of their beat periods (see Eq.~\ref{eq:pbeat}). Both also show periodicity at the stellar rotation period, albeit at a lesser extent than at the aforementioned values. For the sum of the two signals, the periodicity mirrors that of the signal induced by planet b, which is roughly an order of magnitude stronger than that induced by planet c due to its close proximity to the star (see top panel).}
\label{fig:planet signal}
\end{figure}

Over the 500 rotations of the star, we compute the positions of the footpoints of the lines connecting to both planets b and c. If the footpoints lie on the hemisphere visible to the observer (as illustrated in the left panel of Figure~\ref{fig:sketch planet}), we compute the power deposited at the footpoint using Eq.~\ref{eq:planet-induced power}. The resulting simulated power induced by the planets is shown in the top panel of Figure~\ref{fig:planet signal}. The simulated signals from planets b and c are visible 66 and 46\% of the time respectively, with either signal being visible for 81\% of the time. The estimated signal has an average strength of $1.6\times10^{15}$~W, and a peak strength of $1.3\times10^{16}$~W. Note here that we explicitly assume that all of the power carried by the Alfv\'en waves is deposited at the footpoints. In reality however, it is likely that only a small fraction of the energy would be converted into enhanced emission in the chromospheric lines.

In the bottom panel of Figure~\ref{fig:planet signal}, we show the associated GLS periodogram computed from the simulated planet-induced signal. We find that for the individual simulated signals induced by planets b and c, there is strong periodicity at their respective orbital and beat periods, in line with the estimates of \citetalias{fischer2019}. We also see significant periodicity at harmonics of the beat periods, and at the stellar rotation period, albeit to a lesser extent. We note that no such peak can be firmly identified from the periodograms shown in Fig.~\ref{fig:1D_periodogram}. In the same way, no modulation at 8.33\,d is seen from the computed planet-induced power, which might just reflect the fact that the magnetic topology of AU Mic has significantly evolved since 2019. The respective stellar rotation periods required to obtain P$_{\rm{beat}}$\,=\,8.33\,d for \aum\,b and c are 4.2\,d and 5.7\,d, i.e., inconsistent with the equatorial and polar rotation periods of 4.675\,d and 5.34\,d reported in \citet{klein2021}. Similarly, the peak at 8.33\,d does not match the beat period between the two transiting planets nor its first harmonic. As a consequence, the observed 8.33-d modulation, if indeed attributable to a SPMI, is likely a combination of several processes hard to disentangle without simultaneous information on the magnetic topology.

\subsubsection{Comparison to the emitted power of the chromospheric lines}

We now compare the power of the periodic emission lines to that available via sub-Alfv\'enic interactions with planets b and c. Following the methodology introduced in \citet{strugarek2019}, we assume that the continuum spectrum of the chromospheric lines is locally well-described by a blackbody with \teff\,=\,3700\,K. The blackbody radiance is integrated over the width of the He I D3 line in our observed spectra, and the result is integrated over all emission angles in order to quantify a flux hypothetically distributed over the whole stellar disk. We then rescale this flux estimate by the relative area of the footprint at the stellar surface, assumed here to be of 0.1\% \citep{strugarek2019}. We finally fit a double sine-wave at respective periods of 4.8\,d (to account for stellar-induced emission) and 8.33\,d (potential SPMI signature) to the time series of out-of-flare He I emission flux. We find that the emitted power at 8.33\,d features a semi-amplitude of a few $10^{17}$\,W (see Table~\ref{tab:modul_indic}). Similar emitted powers are estimated from the Na I doublet. These values are about one order of magnitude larger than the power estimated from our 3D magnetohydrodynamic simulations (see Fig.~\ref{fig:planet signal}).

The power of the planet-induced signal may be larger than that shown in the top panel of Figure~\ref{fig:planet signal} if planets b/c possess intrinsic magnetic fields. In this scenario, the radius of the planet in Eq.~\ref{eq:planet-induced power} is replaced by the size of the planetary magnetosphere \citep[see][]{kavanagh2021}. The sub-Alfv\'enic region of the stellar wind is dominated by magnetic forces, so the size of the magnetosphere can be approximated as \citep{kavanagh2019}
\begin{eqnarray}
R_\textrm{M} = \Big(\frac{B_\textrm{p}}{B}\Big)^{1/3}~R_\textrm{p},
\label{eq:magnetosphere radius}
\end{eqnarray}
where $B_\textrm{p}$ is the polar magnetic field strength of the planet, and $B$ is the field strength of the stellar wind at the position of the planet. At the orbit of planet b, which induces the majority of the power at the footpoints, the magnetic field strength of the stellar wind is on average 0.1~G. Plugging this value in to Eq.~\ref{eq:magnetosphere radius}, we find a magnetosphere radius of $\sim$5\,\rpb\ to explain the observed emitted power of 3\,$\times$\,10$^{17}$\,W, implying a typical magnetic field of $B_\textrm{p}$\,$\approx$\,10\,G, typically one order of magnitude below the strengths of the magnetospheres of hot Jupiters estimated in \citet{cauley2019}.



An additional source of energy may be produced if the sub-Alfv\'enic star-planet interactions results in the release of energy from coronal loops, as suggested by \citet{lanza2009}. This is thought to occur via the stretching of the footpoint on the stellar surface \citep{lanza2013, cauley2019}. The total power available from this interaction is
\begin{eqnarray}
P = \frac{1}{2} f_\textrm{p} {R_\textrm{p}}^2  {B_\textrm{p}}^2 \Delta u
\label{eq:power_spi2}
\end{eqnarray}
where $f_\textrm{p}$ is the fraction of the planetary hemisphere covered by flux tubes, which depends on the ratio $B/B_\textrm{p}$ \citep{lanza2013}. Assuming that 0.2\% of the total available power is radiated away in He\,I as estimated by \citet{cauley2019} based on Ca II H \& K lines, we would require a planet magnetic field of 10-20\,G to reach the observed power, which is consistent with our estimate from the last paragraph.


We caution that firmly confirming that the observed modulation of the He\,I flux at 8.33\,d is indeed due to a SPMI is a challenging task. First and foremost, the magnetic topology of the star has been demonstrated to be a critical ingredient to hunt for potential signatures of SPMIs. Such information can only be obtained by inverting high-resolution circularly-polarized line profiles into a distribution of the large-scale field vector at the stellar surface with Zeeman-Doppler imaging \citep[see the reviews from][]{donati2009,kochukhov2016}. Hence the need to keep on monitoring the system, but, this time, with high-precision spectropolarimeters (e.g., HARPS-Pol, CRIRES+, SPIRou), in order to simultaneously recover the non-radiative chromospheric emission flux on many lines (spanning both the optical and nIR domains) and the associated magnetic topology. On a final note, we find that including the 6 epochs affected by flares identified in Sec.~\ref{sec:section3.1} tends to increase the significance of the emission flux modulated at 8.33\,d. More spectroscopic observations of consecutive flaring events of \aum\ in different spectral domains (from X-rays to nIR) would greatly help investigating whether some of these flares could be induced by magnetic interactions between the star and its close-in planets, as suggested in \citet{lanza2018} and \citetalias{fischer2019}.

\subsection{New mass estimates for AU Mic close-in planets}

From our estimates of $K_{\rm{b}}$\,=\,7.1\,$\pm$\,3.8\,\ms\ and $K_{\rm{c}}$\,=\,13.3\,$\pm$\,4.1\,\ms, we obtain planet masses of $M_{\rm{b}}$\,=\,14.3\,$\pm$\,7.7\,\mearth\ and $M_{\rm{c}}$\,=\,34.9\,$\pm$\,10.8\,\mearth. These estimates are slightly larger but still consistent within 1~$\sigma$ with the planet masses obtained in \citetalias{zicher2022} by jointly modelling the stellar activity signals and the planet RV signatures using a multi-dimensional Gaussian Process framework \citep{Rajpaul2015,barragan2021}. Over the 121 days spanned by our 2021 data set, \aum's brightness topology has significantly evolved (see Fig.~\ref{fig:prof_fit_aumic_2021}). As this intrinsic evolution is not yet included in our DI code, the corresponding masses are significantly less precise than their counterparts in \citetalias{zicher2022} (see the perspectives of improvement in Sec.~\ref{sec:sect5.4}). Still, we emphasize that the planet mass estimates reported in this study are obtained using a fully independent method to that used in \citetalias{zicher2022}. Therefore, having consistent planet mass estimates between the two studies reinforces the robustness of the planet detection from the HARPS data set.

Similarly to \citetalias{zicher2022}, the masses of \aum\,b and \aum\,c are respectively found slightly lower and significantly larger than their counterparts in \citet{cale2021}, who conducted a velocimetric analysis of a large number of data points collected with multiple optical and nIR spectrographs. The origin of this discrepancy, discussed in more details in the section~4.1 of \citetalias{zicher2022}, remains unclear. Hence the need to (i)~continue the dense RV follow-up of the \aum\ system using different high-precision velocimeters and (ii)~carry out independent analyses led by different teams.

Our mass estimates follow the same trend as those reported in \citetalias{zicher2022}, i.e. \aum\,c, although smaller than planet b, is significantly more massive. Potential planet formation and evolution scenarios are discussed in detail in the Section~4.5 of \citetalias{zicher2022}. One of the main outcomes of this analysis is that the bulk densities of the two planets appear to be in tension with standard core accretion scenarios \citep[e.g.,][]{pollack1996,lee2015}. First, both planets seem to have accreted less H/He from their parent disk than predicted by standard core accretion models. Second, the origin of the high density of planet c is hard to reconcile with planet formation and evolution models given that more massive cores are expected to accrete more H/He in standard core accretion scenarios. These results confirm the crucial need to obtain precise estimates of the mean inner densities of planets transiting very young PMS stars \citep[e.g., K2-33, V1298\,Tau, HIP\,67522 TOI\,1227, reported in][respectively]{David2016a,David2019b,rizzuto2020,mann2021,suarez2021}, in order to resolve potential systematic discrepancies between planet formation and evolution models and the observations.

\subsection{Stellar activity and differential rotation}\label{sec:section5_DR}

The DI process yields respective spot coverages of 4.7 and 5.1\% for the 2020 and 2021 periods, $\sim$3.5 times larger than the 1.4\% obtained from the spectra collected in 2019 with the near-infrared (nIR) spectropolarimeter SPIRou \citep{klein2021}. This increase is almost twice as large as that expected by rescaling our brightness distributions by the Planck's law to account for the decrease in feature contrast between the optical and nIR spectral domains \citep[see Fig.~6 of][]{cale2021}. This excess of spot coverage might either indicate a significant evolution of the stellar activity since 2019 or reflect the fact that the line profile distortions do not trace the same physical phenomena in the nIR, more sensitive to the small-scale field as highlighted in \citet{klein2021}. Once again, spectropolarimetric observations are needed to constrain the strength and topology of the large-scale magnetic field of \aum\ and disentangle between the two scenarios.

We detect a weak DR of \dome\,=\,0.01\,\radd\ shearing the 2021 brightness distribution of \aum\ whereas the 2020 counterpart is consistent with a solid-body rotation. These values are significantly weaker than the DR level of \dome\,=\,0.075\,$\pm$\,0.031\,\radd\ detected from the SPIRou nIR unpolarized line profiles collected in 2019 \citep{klein2021}. We caution that, since most of the features reconstructed with DI are located near the stellar equator (see Fig.~\ref{fig:maps_aumic}), our weak DR estimate might just reflect the partial latitudinal coverage of our brightness maps. We investigate this effect from synthetic CCFs created from the best-fitting brightness distributions shown in Fig.~\ref{fig:maps_aumic}. For each period (i.e., 2020 and 2021), we generate 20 mock CCF time-series assuming different level of DR, from \dome\,=\,0 to \dome\,=\,0.17\,\radd\ \citepalias[corresponding to the level DR detected from the large-scale magnetic field in][]{klein2021}, and inject random noise at the same level as in our observations. In order to assess whether an hypothetical DR would have been detected in our data, we apply our DI code on all these data sets using the method described in Sec.~\ref{subsec:4.1}, and estimate the significance of a model at the best-fitting DR parameters compared to a model assuming a solid-body rotation. This process is then repeated for 10 different realisations of white noise. For both 2020 and 2021 periods, we find that a DR would have been accurately detected at a significance larger than 5$\sigma$. We therefore discard strong biases towards low DR values from our actual DI reconstructions. However, once again, it is difficult to draw conclusions about the DR evolution from the 2019 SPIRou observations of \aum\ and our HARPS observations. As outlined in \citet{klein2021}, the activity-induced distortions of the intensity line profiles observed with SPIRou are most likely of magnetic origin, whereas the variations of HARPS CCFs are primarily due to contrast effects \citep[as expected from the youth and late spectral type of \aum; e.g.,][]{hebrard2014,beeck2015,baroch2020,panja2020}. As a consequence, one alternative explanation for the observed discrepancy in DR is that the brightness distribution probed by HARPS CCFs is anchored at a different depth of the convective zone than the small-scale magnetic field probed by the SPIRou line profiles.

We also note that the time series of chromospheric indices are primarily modulated at rotation periods of typically 4.80\,$\pm$\,0.01\,d (see Sec.~\ref{sec:section2}), which cannot be directly reconciled with the rotation periods unveiled from the observed HARPS CCFs (equatorial periods of 4.85\,$\pm$\,0.01\,d and 4.8559\,$\pm$\,0.0009\,d in 2020 and 2021, respectively). Since the non-radiative chromospheric emission is more sensitive to the surface magnetic field of the star, it reinforces the idea that the small-scale magnetic field probes slightly different layers of the stellar interior in the case of \aum. We note that a contemporary spectropolarimetric monitoring of \aum\ using both optical (e.g., HARPS-Pol, ESPaDOnS, Neo-NARVAL) and nIR (e.g., SPIRou, CRIRES+) should be able to confirm or infirm this assumption. Moreover, the magnetic properties derived from time series of circularly-polarized Zeeman signatures with Zeeman-Doppler imaging have been shown to be excellent proxies of solar-like magnetic cycles \citep{lehmann2021}. Monitoring the long-term evolution of the large-scale field at the surface of \aum\ will allow us to (i)~unveil the origin of the discrepancies in spot coverage and latitudinal DR between HARPS and SPIRou observations and (ii)~help investigate the nature of the putative 5-yr activity cycle reported in \citet{ibanez2019}.


\subsection{Perspectives: filtering stellar activity using Doppler Imaging}\label{sec:sect5.4}

This study confirms that DI has the potential to be used as a reliable independent alternative to the search for planet signatures in RV time series, as it leverages both the spectral and temporal information available in the CCFs to model activity-induced profile distortions whilst leaving planet-induced profile shifts untouched \citep[similarly to what is done in][but with a more physically-motivated approach]{cameron2021}. For the time being, our DI code has been shown to give reliable estimates of close-in planet velocimetric parameters provided that (i)~the star's rotation cycle is densely observed on time scales on which the stellar activity does not significantly evolve, (ii)~the star rotates relatively fast (\vs\,\ga\,10\,\kms), (iii)~the noise level in the observed CCFs is low enough to resolve the profile distortions, and (iv)~the total time span of the observations is significantly larger than the planet orbital period.

However, reliable prospects for overcoming these limitations seem to be emerging. First and foremost, including the intrinsic evolution of the brightness topology, for example, by coupling the DI reconstruction to a stochastic process \citep[as proposed by][and Finociety et al., in prep.]{yu2019,finociety2021,luger2021} will help increase the model flexibility and, thus, will allow one to combine more observations in the DI reconstruction process, thereby obtaining more precise estimates of star/planet parameters. Second, recent efforts to solve the DI inversion in the Bayesian framework suggests that a robust exploration of the parameter space, not dependent on the initial conditions and not affected by local \chisqr\ minimums, might soon be permitted in the DI framework \citep{asensio2021}. Finally, we emphasize that DI can be applied to slow rotators, as already demonstrated in \citet{hebrard2016}, as long as the observations are precise-enough to resolve the activity-induced profile distortions and densely cover the stellar rotation cycle. Systematic differences between the intrinsic profile used in the DI code and the observed CCFs have been shown to be efficiently removed through the iterative process described in \citet{klein2021}.

Under these conditions, it becomes conceivable to apply DI to model the profile distortion of solar-like stars and search for planets around them. However, the activity-induced profile distortions of these stars are known to be primarily affected by the inhibition of convective blueshift in active regions rather than by changes in contrast of the photospheric brightness \citep[e.g.,][]{meunier2010,milbourne2019}. Such effects can be easily incorporated to DI as demonstrated by \citet{hebrard2016}, by redshifting local profiles within regions of strong magnetic field in the reconstruction process. It may therefore be possible to invert a time series of CCFs into surface distributions of relative brightness (spots/plages) and local redshift (i.e., corresponding to regions where the inhibition of convective blueshift is maximum). Since both effects have been shown to affect different spectral lines \citep[e.g.,][]{cretignier2019}, it might be possible to produce two CCF time-series, each primarily affected by one physical process, and to simultaneously invert them into two different maps while searching for the same planet signatures in both time series \citep[as proposed by][]{meunier2017}. The high-resolution solar spectra collected as part as an intensive monitoring campaign with HARPS-N \citep{dumusque2015,cameron2019,dumusque2021} is the ideal benchmark for testing and validating these new methods \citep[e.g., by comparing the resulting maps with data products from the Solar Dynamics Observatory;][]{pesnell2012,Haywood2016,haywood2020}. Applying these methods to other solar-type stars will require to follow them up intensely with state-of-the-art spectrometers as proposed, for example, by the Terra Hunting Experiment\footnote{\url{http://www.terrahunting.org/index.html}} \citep[THE;][]{hall2018}.

\section*{Acknowledgements}

We gratefully acknowledge X. Dumusque and F. Bouchy for their coordination of the HARPS time-share, which allowed us to spread the observations over the two semesters with the appropriate time-sampling, and all the observers involved in the time-share (F. Bouchy,B. Canto, I. de Castro, G. Dransfield, M. Esposito, V. Van Eylen, F.Hawthorn, M. Hobson, V. Hodzic, D. Martin, J. McCormac, H. Os-borne, J. Otegi, A. Suarez, P. Torres) for carrying out the observations on our behalf. We also thank J. Patterson for managing the Oxford Astrophysics compute cluster, glamdring, which was used to carry out the data analysis. This study is based on observations collected at the European Southern Observatory under ESO programmes 0105.C-0288\footnote{The P105 observations were delayed because of the COVID-19 pandemic and taken in P107.} \& 0106.C-0852 (PIs Aigrain and Zicher). This research has made use of NASA’s Astrophysics Data System. B.K., O.B. and S.A. acknowledge funding from the European Research Council under the European Union’s Horizon 2020 research and innovation programme (grant agreement No 865624, GPRV). N.Z. acknowledges support from the UK Science and Technology Facilities Council (STFC) under Grant Code ST/N504233/1, studentship no. 1947725. A.A.V. and R.D.K. acknowledge funding from the European Research Council (ERC) under the European Union's Horizon 2020 research and innovation programme (grant agreement No 817540, ASTROFLOW). L.D.N thanks the Swiss National Science Foundation for support under Early Postdoc.Mobility grant P2GEP2\_200044. A.S. acknowledges PLATO CNES funding at CEA/IFU/DAp and funding from the Programme National de Planétologie (PNP). B.N.’s work is supported by STFC Consolidated Grant ST/S000488/1 (PI Balbus). J.-F.D. and J.B. acknowledge funding from the ERC under the H2020 research and innovation programme (grant agreements No 740651 NewWorlds, and No 742095 SPIDI: Star-Planets-Inner Disk-Interactions). Finally, we thank the reviewer for valuable comments and
suggestions which helped us improving an earlier version of the manuscript.

\section*{Data Availability}

The HARPS spectra used in this study are available on the ESO archive\footnote{\url{http://archive.eso.org/eso/eso_archive_main.html}}.



\bibliographystyle{mnras}
\bibliography{biblio}



\appendix

\section{Journal of observations}

The full journal of observations including the observation dates, noise values in the CCFs, RV/FWHM/BIS time-series, and the chromospheric activity indices computed in Sec.~\ref{sec:section3.1} is given in Table~\ref{tab:journal_obs}.

\begin{table*}
\caption{Journal of HARPS observations analysed in this study. The first 4 columns respectively list the UT date, the BJD at mid-exposure of each observation, the signal-to-noise ratio at 550\,nm, the RMS noise level in the CCF. Column~5 indicates spectra polluted by the Moon or affected by a stellar flare. The time series of DRS-provided RVs, FWHM and BIS are respectively listed in columns~6 to 8 (we adopt a typical error bar of 15\,\ms\ for the FWHM and BIS values). Finally, we give chromospheric emission indices based on Ca\,II H \& K (3968.47 \& 3933.66\,{\AA}), H$\alpha$ (6562.808\,{\AA}), H$\beta$ (4861.363\,{\AA}), Na\,I D1 \& D2 (5895.92 \& 5889.95\,{\AA}) and He\,I D3 (5875.62\,{\AA}) in the last 5 columns.}
\label{tab:journal_obs}
\tiny
\centering
\begin{tabular}{cccccr@{$\pm$}lccr@{$\pm$}lr@{$\pm$}lr@{$\pm$}lr@{$\pm$}lr@{$\pm$}l}
\hline
UT\,Date & BJD  & S/N & $\sigma_{\rm{CCF}}$ & Comments & \multicolumn2c{RV} & FWHM & BIS & \multicolumn2c{S$_{\rm{Ca II}}$} & \multicolumn2c{H$\alpha$} & \multicolumn2c{H$\beta$} & \multicolumn2c{Na\,I} & \multicolumn2c{He\,I}\\
-- & [BJD$_{\rm{TDB}}$ - 2\,457\,000] & -- & [$\times$10$^{-4}$] & -- & \multicolumn2c{[\kms]} & [\kms] & [\kms] & \multicolumn2c{--} & \multicolumn2c{--} & \multicolumn2c{--} & \multicolumn2c{--} & \multicolumn2c{--}\\
\hline
2020-11-15 & 2168.59981 & 75.1 & 4.7 &  & -4.622\, & \,0.007 & 10.54 & -0.16 & 9.2606\, & \,0.0081 & 1.9408\, & \,0.0001 & 2.4122\, & \,0.0004 & 0.2108\, & \,0.0001 & 1.1860\, & \,0.0002 \\
2020-11-17 & 2170.58023 & 59.2 & 5.8 &  & -4.786\, & \,0.009 & 10.78 & -0.02 & 9.8289\, & \,0.0107 & 2.0960\, & \,0.0001 & 2.6658\, & \,0.0006 & 0.2272\, & \,0.0001 & 1.2065\, & \,0.0003 \\
2020-11-18 & 2171.54293 & 73.2 & 4.8 &  & -4.669\, & \,0.007 & 10.73 & -0.03 & 9.6329\, & \,0.0076 & 2.0447\, & \,0.0001 & 2.5515\, & \,0.0004 & 0.2170\, & \,0.0001 & 1.2012\, & \,0.0002 \\
2020-11-19 & 2172.53450 & 108.9 & 3.3 &  & -4.928\, & \,0.005 & 10.71 & 0.09 & 9.5134\, & \,0.0050 & 1.9959\, & \,0.0001 & 2.5150\, & \,0.0002 & 0.2181\, & \,0.0001 & 1.1997\, & \,0.0001 \\
2020-11-22 & 2175.50735 & 101.4 & 3.5 &  & -4.771\, & \,0.005 & 10.76 & -0.03 & 9.7041\, & \,0.0052 & 2.0872\, & \,0.0001 & 2.6374\, & \,0.0003 & 0.2253\, & \,0.0001 & 1.2074\, & \,0.0001 \\
2020-11-22 & 2175.56221 & 84.9 & 4.2 &  & -4.773\, & \,0.006 & 10.80 & -0.05 & 9.9363\, & \,0.0072 & 2.1216\, & \,0.0001 & 2.7238\, & \,0.0003 & 0.2313\, & \,0.0001 & 1.2220\, & \,0.0002 \\
2020-11-23 & 2176.50801 & 99.4 & 3.6 &  & -4.671\, & \,0.005 & 10.77 & -0.03 & 9.7568\, & \,0.0054 & 2.1079\, & \,0.0001 & 2.6359\, & \,0.0003 & 0.2243\, & \,0.0001 & 1.2035\, & \,0.0001 \\
2020-11-24 & 2177.51177 & 107.9 & 3.4 &  & -4.964\, & \,0.005 & 10.67 & 0.12 & 9.7824\, & \,0.0053 & 2.0319\, & \,0.0001 & 2.6041\, & \,0.0002 & 0.2263\, & \,0.0001 & 1.2027\, & \,0.0001 \\
2020-11-25 & 2178.50720 & 93.0 & 3.9 &  & -4.623\, & \,0.006 & 10.67 & -0.17 & 10.1740\, & \,0.0059 & 2.1333\, & \,0.0001 & 2.7701\, & \,0.0003 & 0.2353\, & \,0.0001 & 1.2170\, & \,0.0001 \\
2020-11-25 & 2178.51843 & 88.2 & 4.0 &  & -4.629\, & \,0.006 & 10.69 & -0.15 & 10.2684\, & \,0.0065 & 2.1341\, & \,0.0001 & 2.7660\, & \,0.0003 & 0.2348\, & \,0.0001 & 1.2165\, & \,0.0002 \\
2020-11-26 & 2179.51743 & 114.1 &  -- & Flare & -4.762\, & \,0.005 & 10.79 & -0.07 & 12.9278\, & \,0.0062 & 2.4398\, & \,0.0001 & 3.3473\, & \,0.0002 & 0.2927\, & \,0.0001 & 1.2704\, & \,0.0001 \\
2020-11-27 & 2180.54004 & 107.1 & 3.4 &  & -4.749\, & \,0.005 & 10.81 & -0.02 & 11.0421\, & \,0.0062 & 2.2608\, & \,0.0001 & 2.9794\, & \,0.0003 & 0.2526\, & \,0.0001 & 1.2179\, & \,0.0001 \\
2020-11-28 & 2181.54053 & 86.0 & 4.2 &  & -4.732\, & \,0.006 & 10.83 & 0.02 & 11.4547\, & \,0.0074 & 2.3659\, & \,0.0001 & 3.1749\, & \,0.0004 & 0.2637\, & \,0.0001 & 1.2759\, & \,0.0002 \\
2020-11-29 & 2182.53789 & 108.6 & 3.4 &  & -4.923\, & \,0.005 & 10.53 & 0.14 & 9.7686\, & \,0.0059 & 1.9993\, & \,0.0001 & 2.5468\, & \,0.0003 & 0.2206\, & \,0.0001 & 1.1954\, & \,0.0001 \\
2020-11-30 & 2183.54048 & 123.3 & 3.0 &  & -4.657\, & \,0.004 & 10.77 & -0.07 & 10.7131\, & \,0.0049 & 2.2642\, & \,0.0001 & 2.9157\, & \,0.0002 & 0.2485\, & \,0.0001 & 1.2664\, & \,0.0001 \\
2020-12-01 & 2184.54177 & 126.7 & 2.9 &  & -4.813\, & \,0.004 & 10.83 & -0.00 & 10.3151\, & \,0.0048 & 2.2843\, & \,0.0001 & 2.8855\, & \,0.0002 & 0.2356\, & \,0.0001 & 1.2284\, & \,0.0001 \\
2020-12-02 & 2185.54211 & 73.8 & 4.8 &  & -4.727\, & \,0.007 & 10.78 & 0.01 & 10.0738\, & \,0.0083 & 2.1483\, & \,0.0001 & 2.7128\, & \,0.0004 & 0.2280\, & \,0.0001 & 1.2093\, & \,0.0002 \\
2020-12-05 & 2188.54816 & 95.4 & 3.8 &  & -4.716\, & \,0.005 & 10.71 & -0.02 & 9.4786\, & \,0.0055 & 2.0197\, & \,0.0001 & 2.5273\, & \,0.0003 & 0.2157\, & \,0.0001 & 1.1948\, & \,0.0001 \\
2020-12-06 & 2189.54737 & 122.6 & 3.0 &  & -4.861\, & \,0.004 & 10.81 & 0.07 & 11.5011\, & \,0.0048 & 2.3254\, & \,0.0001 & 3.1736\, & \,0.0002 & 0.2676\, & \,0.0001 & 1.2666\, & \,0.0001 \\
2020-12-07 & 2190.52211 & 110.7 & 3.3 &  & -4.720\, & \,0.005 & 10.79 & 0.07 & 10.8230\, & \,0.0047 & 2.2854\, & \,0.0001 & 3.0350\, & \,0.0002 & 0.2484\, & \,0.0001 & 1.2706\, & \,0.0001 \\
2020-12-08 & 2191.52198 & 128.9 & 2.9 &  & -4.834\, & \,0.004 & 10.81 & 0.10 & 10.4088\, & \,0.0049 & 2.1643\, & \,0.0001 & 2.8234\, & \,0.0002 & 0.2331\, & \,0.0001 & 1.2213\, & \,0.0001 \\
2020-12-09 & 2192.52110 & 128.5 & 2.9 &  & -4.808\, & \,0.004 & 10.41 & 0.03 & 10.2756\, & \,0.0046 & 2.1205\, & \,0.0001 & 2.6961\, & \,0.0002 & 0.2320\, & \,0.0001 & 1.1999\, & \,0.0001 \\
2021-05-24 & 2358.80511 & 138.8 & 2.9 &  & -4.560\, & \,0.004 & 10.74 & -0.18 & 10.0506\, & \,0.0030 & 2.0604\, & \,0.0001 & 2.7077\, & \,0.0002 & 0.2389\, & \,0.0001 & 1.2207\, & \,0.0001 \\
2021-05-25 & 2359.75665 & 120.3 & 3.3 &  & -4.792\, & \,0.004 & 10.52 & 0.02 & 9.3572\, & \,0.0038 & 2.0587\, & \,0.0001 & 2.5386\, & \,0.0002 & 0.2239\, & \,0.0001 & 1.2071\, & \,0.0001 \\
2021-05-25 & 2359.87440 & 122.8 & 3.2 &  & -4.714\, & \,0.004 & 10.58 & -0.06 & 9.2266\, & \,0.0031 & 2.0251\, & \,0.0001 & 2.4693\, & \,0.0002 & 0.2203\, & \,0.0001 & 1.1951\, & \,0.0001 \\
2021-05-26 & 2360.76126 & 115.6 & 3.4 &  & -4.714\, & \,0.005 & 10.86 & -0.01 & 9.5844\, & \,0.0039 & 2.1051\, & \,0.0001 & 2.6118\, & \,0.0002 & 0.2266\, & \,0.0001 & 1.2040\, & \,0.0001 \\
2021-05-26 & 2360.88080 & 95.4 & 4.0 &  & -4.756\, & \,0.005 & 10.84 & 0.04 & 9.8530\, & \,0.0044 & 2.1402\, & \,0.0001 & 2.6966\, & \,0.0003 & 0.2343\, & \,0.0001 & 1.2188\, & \,0.0001 \\
2021-05-27 & 2361.78805 & 84.8 & 4.6 &  & -4.943\, & \,0.006 & 10.71 & 0.13 & 9.7470\, & \,0.0053 & 2.1024\, & \,0.0001 & 2.6250\, & \,0.0004 & 0.2292\, & \,0.0001 & 1.2214\, & \,0.0002 \\
2021-06-01 & 2366.73319 & 111.0 & 3.5 &  & -4.978\, & \,0.005 & 10.68 & 0.12 & 9.7278\, & \,0.0041 & 2.0791\, & \,0.0001 & 2.5945\, & \,0.0002 & 0.2251\, & \,0.0001 & 1.2095\, & \,0.0001 \\
2021-06-01 & 2366.85758 & 86.2 & 4.3 &  & -4.944\, & \,0.006 & 10.64 & 0.07 & 10.4283\, & \,0.0047 & 2.1398\, & \,0.0001 & 2.7547\, & \,0.0003 & 0.2366\, & \,0.0001 & 1.2240\, & \,0.0002 \\
2021-06-02 & 2367.73664 & 97.9 & 4.0 &  & -4.857\, & \,0.005 & 10.63 & 0.10 & 10.7515\, & \,0.0050 & 2.1716\, & \,0.0001 & 2.8956\, & \,0.0003 & 0.2548\, & \,0.0001 & 1.2720\, & \,0.0001 \\
2021-06-02 & 2367.84344 & 87.9 & 4.3 &  & -4.784\, & \,0.006 & 10.52 & 0.04 & 10.9900\, & \,0.0054 & 2.2166\, & \,0.0001 & 2.9121\, & \,0.0003 & 0.2600\, & \,0.0001 & 1.2682\, & \,0.0001 \\
2021-06-03 & 2368.81209 & 110.8 & 3.5 &  & -4.711\, & \,0.005 & 10.99 & -0.02 & 9.0297\, & \,0.0037 & 1.9376\, & \,0.0001 & 2.3695\, & \,0.0002 & 0.2153\, & \,0.0001 & 1.1890\, & \,0.0001 \\
2021-06-03 & 2368.91402 & 107.5 & 3.7 &  & -4.785\, & \,0.005 & 10.95 & 0.04 & 9.2627\, & \,0.0049 & 1.9917\, & \,0.0001 & 2.5019\, & \,0.0003 & 0.2229\, & \,0.0001 & 1.2151\, & \,0.0001 \\
2021-06-04 & 2369.79091 & 131.8 & 2.9 &  & -4.615\, & \,0.004 & 10.70 & -0.13 & 9.9850\, & \,0.0030 & 2.0953\, & \,0.0001 & 2.6691\, & \,0.0002 & 0.2319\, & \,0.0001 & 1.2182\, & \,0.0001 \\
2021-06-04 & 2369.92061 & 120.4 & 3.3 &  & -4.603\, & \,0.004 & 10.79 & -0.10 & 9.3959\, & \,0.0043 & 2.0505\, & \,0.0001 & 2.5908\, & \,0.0002 & 0.2218\, & \,0.0001 & 1.2038\, & \,0.0001 \\
2021-06-05 & 2370.76677 & 83.5 & 4.6 &  & -4.774\, & \,0.006 & 10.82 & 0.01 & 9.9179\, & \,0.0052 & 2.1579\, & \,0.0001 & 2.7223\, & \,0.0004 & 0.2308\, & \,0.0001 & 1.2124\, & \,0.0002 \\
2021-06-06 & 2371.78083 & 124.1 &  -- & Flare & -4.930\, & \,0.004 & 10.71 & 0.06 & 12.5039\, & \,0.0039 & 2.5356\, & \,0.0001 & 3.5115\, & \,0.0002 & 0.2844\, & \,0.0001 & 1.3054\, & \,0.0001 \\
2021-06-06 & 2371.90653 & 113.8 & 3.6 &  & -4.921\, & \,0.005 & 10.65 & 0.06 & 10.4133\, & \,0.0050 & 2.2920\, & \,0.0001 & 2.9205\, & \,0.0003 & 0.2420\, & \,0.0001 & 1.2244\, & \,0.0001 \\
2021-06-07 & 2372.75324 & 92.7 & 4.2 &  & -4.778\, & \,0.006 & 10.48 & 0.05 & 9.7121\, & \,0.0048 & 2.0316\, & \,0.0001 & 2.5870\, & \,0.0003 & 0.2317\, & \,0.0001 & 1.2100\, & \,0.0001 \\
2021-06-11 & 2376.72259 & 108.6 & 3.6 &  & -4.922\, & \,0.005 & 10.62 & 0.05 & 10.6094\, & \,0.0044 & 2.1554\, & \,0.0001 & 2.7781\, & \,0.0003 & 0.2423\, & \,0.0001 & 1.2192\, & \,0.0001 \\
2021-06-12 & 2377.74066 & 120.9 & 3.2 &  & -4.711\, & \,0.004 & 10.44 & 0.01 & 10.8036\, & \,0.0038 & 2.1599\, & \,0.0001 & 2.7811\, & \,0.0002 & 0.2509\, & \,0.0001 & 1.2341\, & \,0.0001 \\
2021-06-13 & 2378.78840 & 115.3 & 3.4 &  & -4.859\, & \,0.005 & 10.82 & 0.12 & 9.7233\, & \,0.0036 & 1.9942\, & \,0.0001 & 2.4764\, & \,0.0002 & 0.2237\, & \,0.0001 & 1.2014\, & \,0.0001 \\
2021-06-14 & 2379.86158 & 90.8 & 4.3 &  & -4.652\, & \,0.006 & 10.82 & -0.08 & 10.2452\, & \,0.0057 & 2.2032\, & \,0.0001 & 2.9424\, & \,0.0003 & 0.2517\, & \,0.0001 & 1.2620\, & \,0.0001 \\
2021-07-10 & 2405.78455 & 122.2 & 3.2 &  & -4.968\, & \,0.004 & 10.56 & 0.08 & 11.0842\, & \,0.0041 & 2.2592\, & \,0.0001 & 2.9694\, & \,0.0002 & 0.2577\, & \,0.0001 & 1.2597\, & \,0.0001 \\
2021-07-11 & 2406.71022 & 72.1 & 5.3 &  & -4.815\, & \,0.007 & 10.54 & 0.12 & 9.5724\, & \,0.0065 & 1.9712\, & \,0.0001 & 2.4489\, & \,0.0004 & 0.2229\, & \,0.0001 & 1.1952\, & \,0.0002 \\
2021-07-13 & 2408.90951 & 66.2 & 5.9 &  & -4.611\, & \,0.008 & 10.79 & -0.13 & 9.6850\, & \,0.0101 & 2.0544\, & \,0.0001 & 2.5845\, & \,0.0006 & 0.2252\, & \,0.0001 & 1.2019\, & \,0.0002 \\
2021-07-14 & 2409.71290 & 81.4 & 4.8 &  & -4.778\, & \,0.007 & 10.82 & -0.05 & 10.8055\, & \,0.0062 & 2.2442\, & \,0.0001 & 2.9478\, & \,0.0004 & 0.2567\, & \,0.0001 & 1.2479\, & \,0.0002 \\
2021-07-15 & 2410.70717 & 93.6 & 4.1 &  & -4.962\, & \,0.005 & 10.60 & 0.05 & 10.1279\, & \,0.0046 & 2.0802\, & \,0.0001 & 2.6175\, & \,0.0003 & 0.2346\, & \,0.0001 & 1.2121\, & \,0.0001 \\
2021-07-15 & 2410.79291 & 93.7 & 4.2 &  & -4.931\, & \,0.006 & 10.62 & 0.02 & 11.0440\, & \,0.0061 & 2.1888\, & \,0.0001 & 2.9409\, & \,0.0003 & 0.2580\, & \,0.0001 & 1.2366\, & \,0.0001 \\
2021-07-16 & 2411.74121 & 121.3 & 3.2 &  & -4.697\, & \,0.004 & 10.45 & 0.02 & 9.1095\, & \,0.0030 & 1.9407\, & \,0.0001 & 2.3077\, & \,0.0002 & 0.2139\, & \,0.0001 & 1.1811\, & \,0.0001 \\
2021-07-17 & 2412.62503 & 101.8 & 3.8 &  & -4.697\, & \,0.005 & 10.96 & -0.01 & 10.0047\, & \,0.0043 & 2.0583\, & \,0.0001 & 2.5822\, & \,0.0003 & 0.2287\, & \,0.0001 & 1.2244\, & \,0.0001 \\
2021-07-17 & 2412.74029 & 92.4 & 4.1 &  & -4.762\, & \,0.006 & 10.90 & 0.05 & 10.0587\, & \,0.0046 & 2.0736\, & \,0.0001 & 2.6265\, & \,0.0003 & 0.2368\, & \,0.0001 & 1.2263\, & \,0.0001 \\
2021-07-27 & 2422.69852 & 78.7 & 4.9 &  & -4.859\, & \,0.007 & 10.71 & 0.08 & 11.1785\, & \,0.0070 & 2.2552\, & \,0.0001 & 3.0334\, & \,0.0004 & 0.2619\, & \,0.0001 & 1.2826\, & \,0.0002 \\
2021-07-27 & 2422.83670 & 72.3 & 5.5 &  & -4.834\, & \,0.007 & 10.63 & 0.03 & 11.1212\, & \,0.0093 & 2.3507\, & \,0.0001 & 3.1260\, & \,0.0005 & 0.2685\, & \,0.0001 & 1.2889\, & \,0.0002 \\
2021-07-28 & 2423.70035 & 71.0 &  -- & Flare & -4.721\, & \,0.008 & 10.95 & -0.02 & 15.1729\, & \,0.0082 & 2.9270\, & \,0.0002 & 4.0772\, & \,0.0005 & 0.4098\, & \,0.0001 & 1.5472\, & \,0.0002 \\
2021-07-28 & 2423.79786 & 76.6 &  -- & Flare & -4.756\, & \,0.007 & 10.90 & 0.05 & 13.8001\, & \,0.0092 & 2.6731\, & \,0.0001 & 3.7837\, & \,0.0005 & 0.3484\, & \,0.0001 & 1.3922\, & \,0.0002 \\
2021-07-29 & 2424.73029 & 117.2 & 3.3 &  & -4.969\, & \,0.005 & 10.92 & 0.12 & 10.6437\, & \,0.0038 & 2.2041\, & \,0.0001 & 2.7893\, & \,0.0002 & 0.2449\, & \,0.0001 & 1.2269\, & \,0.0001 \\
2021-07-30 & 2425.66184 & 92.9 & 4.1 &  & -4.874\, & \,0.006 & 10.67 & 0.01 & 9.9614\, & \,0.0047 & 2.0396\, & \,0.0001 & 2.5177\, & \,0.0003 & 0.2289\, & \,0.0001 & 1.1981\, & \,0.0001 \\
2021-07-31 & 2426.61526 & 80.8 & 4.6 &  & -4.534\, & \,0.006 & 10.48 & -0.11 & 10.0349\, & \,0.0053 & 2.0393\, & \,0.0001 & 2.5706\, & \,0.0004 & 0.2301\, & \,0.0001 & 1.2145\, & \,0.0002 \\
2021-08-01 & 2427.77361 & 122.4 & 3.2 &  & -4.818\, & \,0.004 & 10.66 & 0.05 & 9.4127\, & \,0.0039 & 2.0613\, & \,0.0001 & 2.5355\, & \,0.0002 & 0.2202\, & \,0.0001 & 1.1958\, & \,0.0001 \\
2021-08-02 & 2428.82886 & 106.6 &  -- & Flare & -4.831\, & \,0.005 & 10.78 & 0.13 & 14.0563\, & \,0.0068 & 2.6252\, & \,0.0001 & 3.7482\, & \,0.0003 & 0.3232\, & \,0.0001 & 1.3275\, & \,0.0001 \\
2021-08-03 & 2429.75989 & 90.0 & 4.3 &  & -5.034\, & \,0.006 & 10.78 & 0.20 & 10.5755\, & \,0.0055 & 2.1168\, & \,0.0001 & 2.7080\, & \,0.0003 & 0.2332\, & \,0.0001 & 1.2077\, & \,0.0001 \\
2021-08-04 & 2430.72423 & 44.5 & 8.3 &  & -4.821\, & \,0.012 & 10.73 & 0.04 & 10.5992\, & \,0.0159 & 2.0478\, & \,0.0002 & 2.5133\, & \,0.0008 & 0.2267\, & \,0.0001 & 1.1977\, & \,0.0004 \\
2021-08-06 & 2432.65773 & 88.6 & 4.3 &  & -4.822\, & \,0.006 & 10.64 & 0.03 & 9.8686\, & \,0.0051 & 2.1007\, & \,0.0001 & 2.5649\, & \,0.0003 & 0.2274\, & \,0.0001 & 1.2113\, & \,0.0001 \\
2021-08-07 & 2433.70690 & 97.3 & 4.0 &  & -4.748\, & \,0.005 & 10.84 & 0.06 & 9.8566\, & \,0.0049 & 2.0708\, & \,0.0001 & 2.5987\, & \,0.0003 & 0.2277\, & \,0.0001 & 1.2163\, & \,0.0001 \\
2021-08-16 & 2442.58811 & 77.1 & 5.0 &  & -4.774\, & \,0.007 & 10.67 & 0.03 & 11.3262\, & \,0.0070 & 2.3224\, & \,0.0001 & 3.0185\, & \,0.0004 & 0.2538\, & \,0.0001 & 1.2540\, & \,0.0002 \\
2021-08-17 & 2443.55645 & 118.8 & 3.3 &  & -4.764\, & \,0.004 & 10.81 & 0.05 & 10.8298\, & \,0.0038 & 2.2909\, & \,0.0001 & 2.9997\, & \,0.0002 & 0.2432\, & \,0.0001 & 1.2430\, & \,0.0001 \\
2021-08-17 & 2443.76879 & 123.7 & 3.3 &  & -4.822\, & \,0.004 & 10.84 & -0.03 & 10.5793\, & \,0.0047 & 2.2371\, & \,0.0001 & 2.9192\, & \,0.0002 & 0.2419\, & \,0.0001 & 1.2420\, & \,0.0001 \\
2021-08-18 & 2444.57172 & 127.9 & 3.0 &  & -5.005\, & \,0.004 & 10.58 & 0.12 & 11.0058\, & \,0.0034 & 2.1936\, & \,0.0001 & 2.8456\, & \,0.0002 & 0.2472\, & \,0.0001 & 1.2112\, & \,0.0001 \\
2021-08-18 & 2444.66706 & 128.1 & 3.0 &  & -4.956\, & \,0.004 & 10.54 & 0.06 & 11.7625\, & \,0.0033 & 2.2827\, & \,0.0001 & 3.0541\, & \,0.0002 & 0.2671\, & \,0.0001 & 1.2569\, & \,0.0001 \\
2021-08-18 & 2444.79139 & 128.9 & 3.2 &  & -4.900\, & \,0.004 & 10.57 & 0.02 & 11.0667\, & \,0.0048 & 2.2218\, & \,0.0001 & 2.8983\, & \,0.0002 & 0.2518\, & \,0.0001 & 1.2196\, & \,0.0001 \\
2021-08-23 & 2449.53572 & 87.2 &  -- & Moon & -4.964\, & \,0.006 & 10.53 & 0.06 & 11.1271\, & \,0.0057 & 2.3569\, & \,0.0001 & 2.9417\, & \,0.0003 & 0.2337\, & \,0.0001 & 1.2707\, & \,0.0001 \\
2021-08-23 & 2449.65050 & 64.1 &  -- & Moon & -4.889\, & \,0.008 & 10.57 & 0.01 & 10.6075\, & \,0.0082 & 2.2170\, & \,0.0002 & 2.7339\, & \,0.0005 & 0.2301\, & \,0.0001 & 1.2280\, & \,0.0002 \\
2021-08-23 & 2449.78767 & 44.3 &  -- & Moon & -4.909\, & \,0.012 & 10.67 & -0.04 & 9.4029\, & \,0.0153 & 2.1991\, & \,0.0002 & 2.6674\, & \,0.0009 & 0.2313\, & \,0.0001 & 1.2316\, & \,0.0004 \\
2021-09-03 & 2460.65065 & 131.4 & 3.0 &  & -4.556\, & \,0.004 & 10.48 & -0.10 & 9.5292\, & \,0.0035 & 1.9627\, & \,0.0001 & 2.3957\, & \,0.0002 & 0.2178\, & \,0.0001 & 1.1853\, & \,0.0001 \\
2021-09-04 & 2461.56341 & 113.4 & 3.4 &  & -4.823\, & \,0.005 & 10.84 & 0.06 & 9.8331\, & \,0.0036 & 2.0288\, & \,0.0001 & 2.5706\, & \,0.0002 & 0.2266\, & \,0.0001 & 1.2025\, & \,0.0001 \\
2021-09-05 & 2462.73246 & 92.7 & 4.3 &  & -4.717\, & \,0.006 & 10.88 & 0.05 & 11.8320\, & \,0.0066 & 2.3776\, & \,0.0001 & 3.1975\, & \,0.0003 & 0.2731\, & \,0.0001 & 1.2613\, & \,0.0001 \\
2021-09-06 & 2463.78569 & 97.8 & 4.1 &  & -5.071\, & \,0.006 & 10.69 & 0.20 & 10.3435\, & \,0.0065 & 2.1438\, & \,0.0001 & 2.7573\, & \,0.0003 & 0.2428\, & \,0.0001 & 1.2165\, & \,0.0001 \\
2021-09-07 & 2464.53267 & 106.7 & 3.7 &  & -4.821\, & \,0.005 & 10.71 & -0.01 & 11.7331\, & \,0.0047 & 2.2921\, & \,0.0001 & 3.0614\, & \,0.0003 & 0.2719\, & \,0.0001 & 1.2176\, & \,0.0001 \\
2021-09-10 & 2467.59928 & 136.5 & 2.8 &  & -4.688\, & \,0.004 & 10.89 & -0.00 & 9.6371\, & \,0.0024 & 2.0236\, & \,0.0001 & 2.5127\, & \,0.0002 & 0.2263\, & \,0.0001 & 1.2078\, & \,0.0001 \\
2021-09-13 & 2470.60654 & 95.3 & 4.2 &  & -4.464\, & \,0.006 & 10.63 & -0.21 & 9.1327\, & \,0.0050 & 1.9328\, & \,0.0001 & 2.3523\, & \,0.0003 & 0.2144\, & \,0.0001 & 1.1863\, & \,0.0001 \\
2021-09-14 & 2471.55999 & 135.7 & 2.9 &  & -4.792\, & \,0.004 & 10.64 & 0.03 & 10.8671\, & \,0.0030 & 2.2237\, & \,0.0001 & 2.9136\, & \,0.0002 & 0.2547\, & \,0.0001 & 1.2740\, & \,0.0001 \\
2021-09-15 & 2472.57476 & 117.1 &  -- & Flare & -4.785\, & \,0.005 & 10.82 & 0.03 & 10.8393\, & \,0.0031 & 2.6923\, & \,0.0001 & 3.4396\, & \,0.0002 & 0.3142\, & \,0.0001 & 1.3673\, & \,0.0001 \\
2021-09-18 & 2476.49416 & 108.9 & 3.6 &  & -4.839\, & \,0.005 & 10.63 & 0.05 & 10.0083\, & \,0.0039 & 2.1494\, & \,0.0001 & 2.6948\, & \,0.0002 & 0.2362\, & \,0.0001 & 1.2195\, & \,0.0001 \\
2021-09-20 & 2477.52519 & 70.9 & 5.4 &  & -4.753\, & \,0.008 & 10.89 & -0.01 & 9.7041\, & \,0.0072 & 2.0953\, & \,0.0001 & 2.6260\, & \,0.0005 & 0.2308\, & \,0.0001 & 1.2084\, & \,0.0002 \\
2021-09-20 & 2477.63321 & 62.0 & 6.3 &  & -4.802\, & \,0.009 & 10.92 & -0.01 & 9.2273\, & \,0.0092 & 2.0483\, & \,0.0002 & 2.5155\, & \,0.0006 & 0.2239\, & \,0.0001 & 1.1899\, & \,0.0002 \\
2021-09-20 & 2478.50883 & 96.4 & 4.0 &  & -5.040\, & \,0.005 & 10.59 & 0.17 & 10.4409\, & \,0.0045 & 2.1095\, & \,0.0001 & 2.6825\, & \,0.0003 & 0.2380\, & \,0.0001 & 1.2080\, & \,0.0001 \\
2021-09-21 & 2478.65930 & 110.9 & 3.6 &  & -4.976\, & \,0.005 & 10.57 & 0.10 & 11.0016\, & \,0.0049 & 2.1774\, & \,0.0001 & 2.8682\, & \,0.0003 & 0.2571\, & \,0.0001 & 1.2491\, & \,0.0001 \\
2021-09-22 & 2479.51606 & 54.5 & 6.8 &  & -4.759\, & \,0.009 & 10.71 & 0.06 & 9.9181\, & \,0.0090 & 2.0058\, & \,0.0002 & 2.4825\, & \,0.0006 & 0.2270\, & \,0.0001 & 1.2233\, & \,0.0003 \\
2021-09-22 & 2479.71843 & 85.8 & 4.7 &  & -4.770\, & \,0.006 & 10.53 & 0.08 & 10.3132\, & \,0.0072 & 2.1107\, & \,0.0001 & 2.7327\, & \,0.0004 & 0.2404\, & \,0.0001 & 1.2201\, & \,0.0001 \\
\hline
\end{tabular}
\end{table*}

\section{Activity proxies}

\subsection{Chromospheric emission lines}

\begin{figure*}
\begin{center}
        \includegraphics[width=0.6\textwidth]{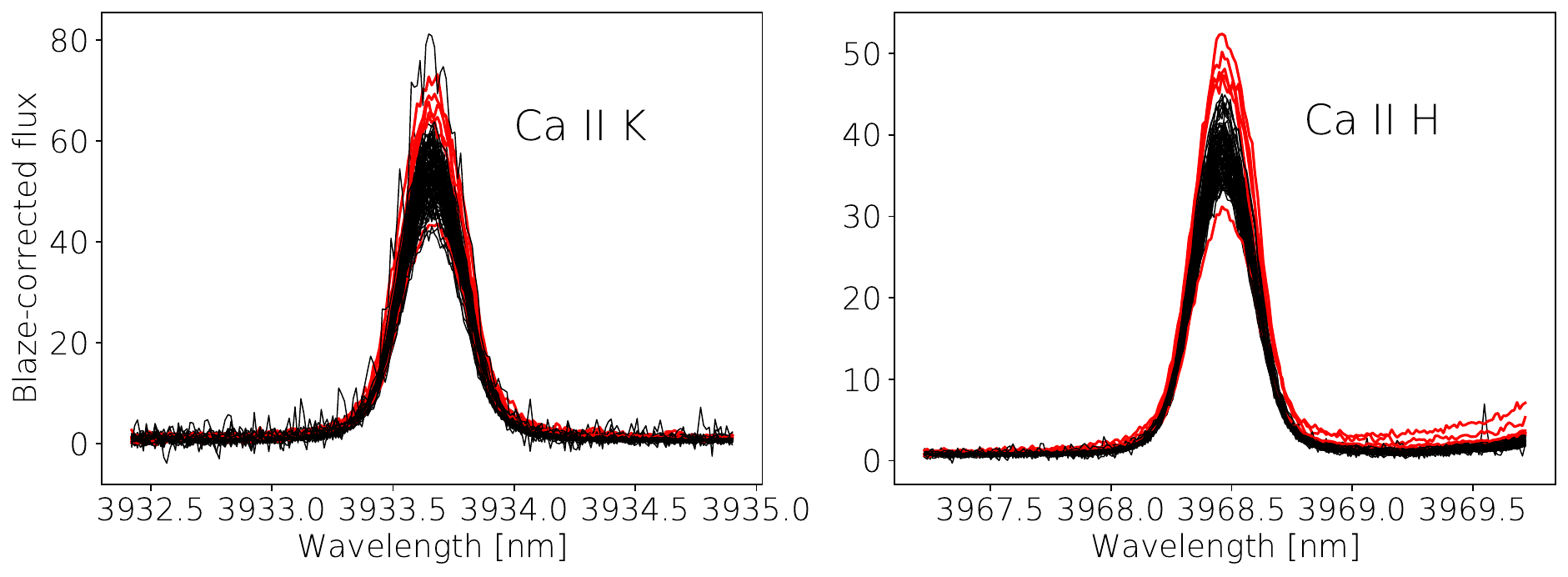}\ 
        \includegraphics[width=0.3\textwidth]{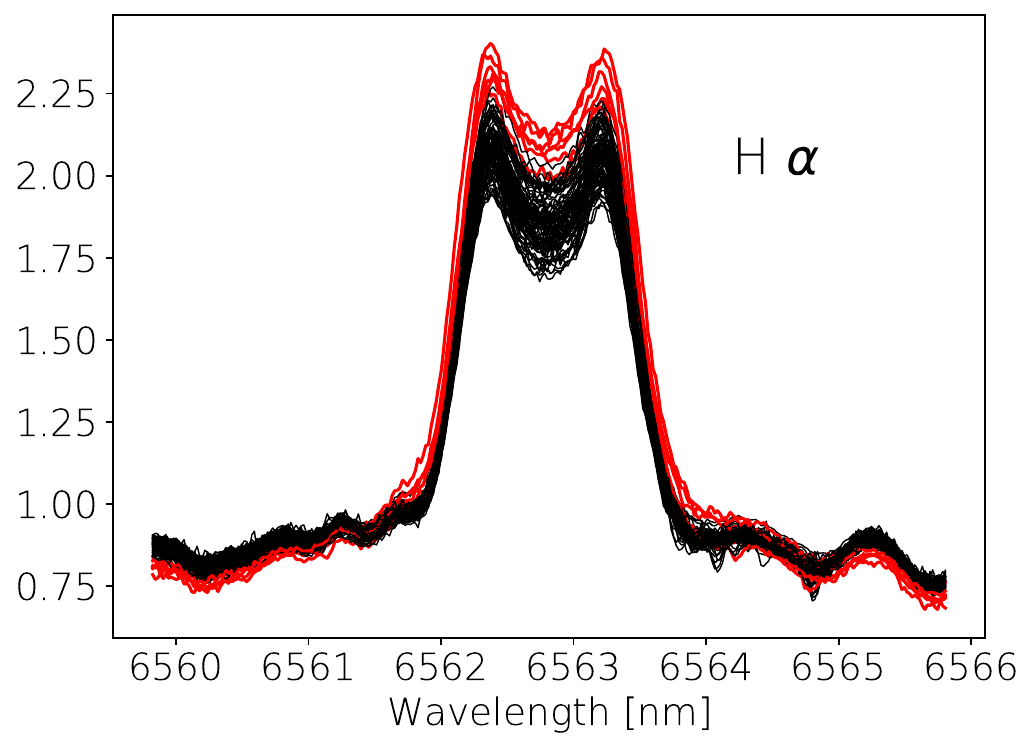}\\ 
        \includegraphics[width=0.6\textwidth]{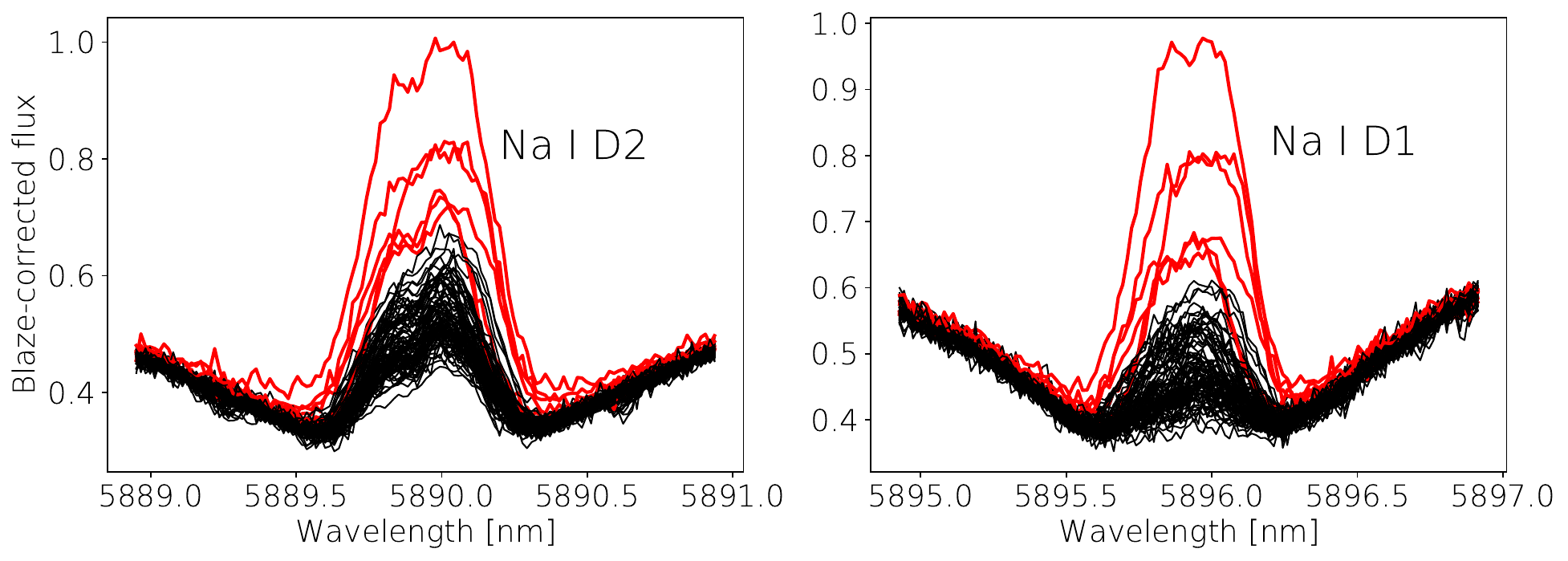}\ 
        \includegraphics[width=0.3\textwidth]{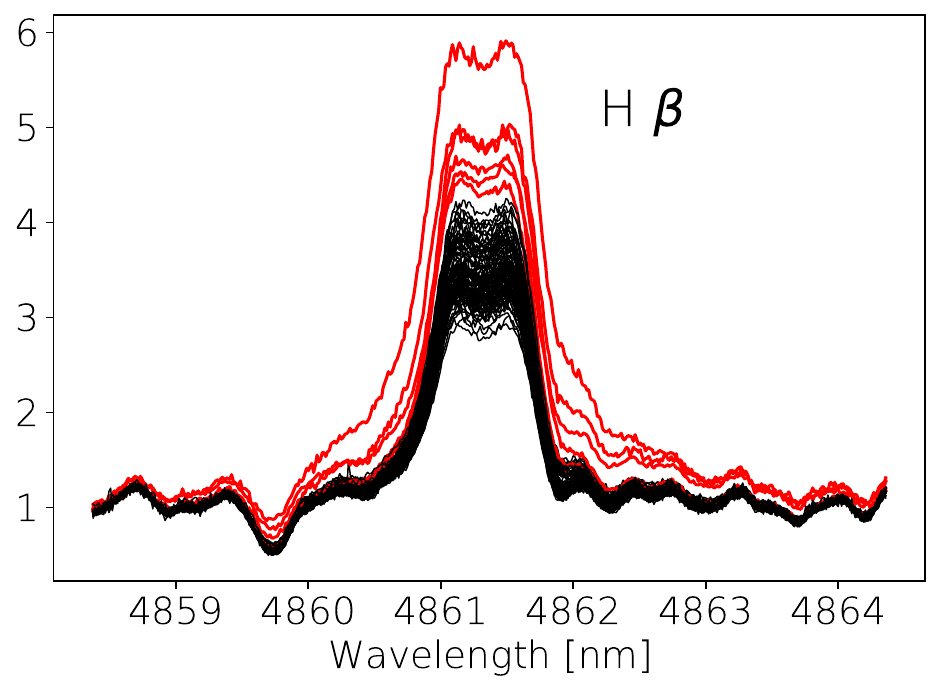}\\ 
        \includegraphics[width=0.3\textwidth]{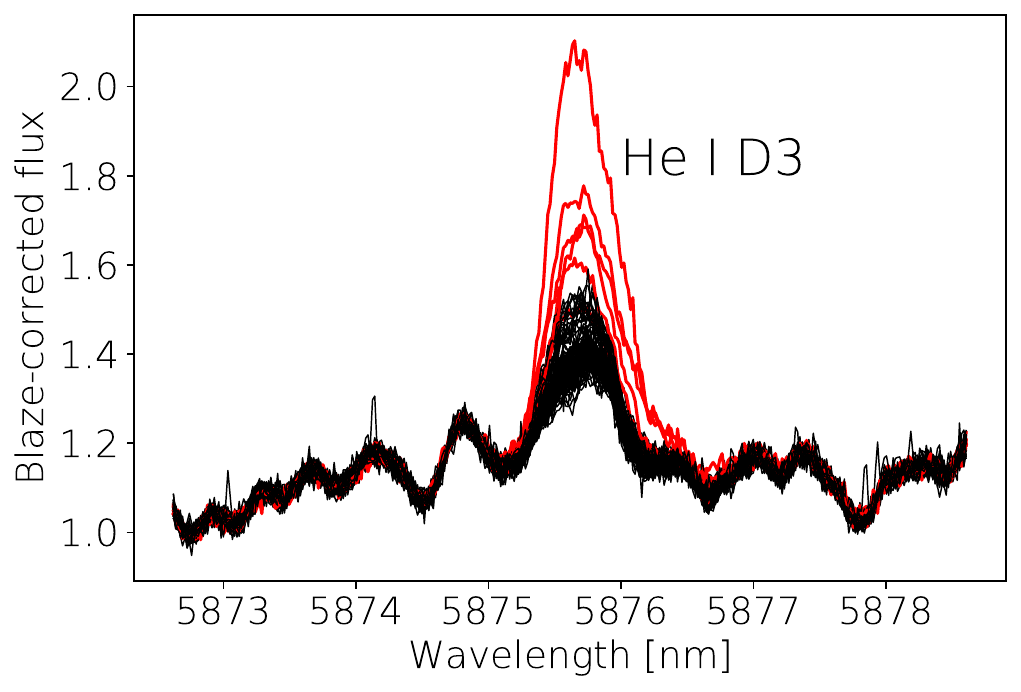} 
\caption{Observed emission reversals of Ca II H \& K, H$\alpha$, Na I D1 \& D2, H$\beta$ and He I D3 chromospheric lines for \aum. In each panel, profiles shown in red were flagged as affected by a stellar flare (6 observations). Note that the Moon-polluted observations (2021--08--23) are not included in these plots.}
\label{fig:detail_chromospheric_lines}
\end{center}
\end{figure*}

The observed Ca\,II H \& K, H$\alpha$, H$\beta$, Na\,I D1 \& D2 and He\,I D3 profiles are shown in units of blaze-corrected flux in the stellar rest frame in Fig.~\ref{fig:detail_chromospheric_lines}. 

\subsection{Correlation analysis}

\begin{figure*}
    \centering
    \includegraphics[width=\linewidth]{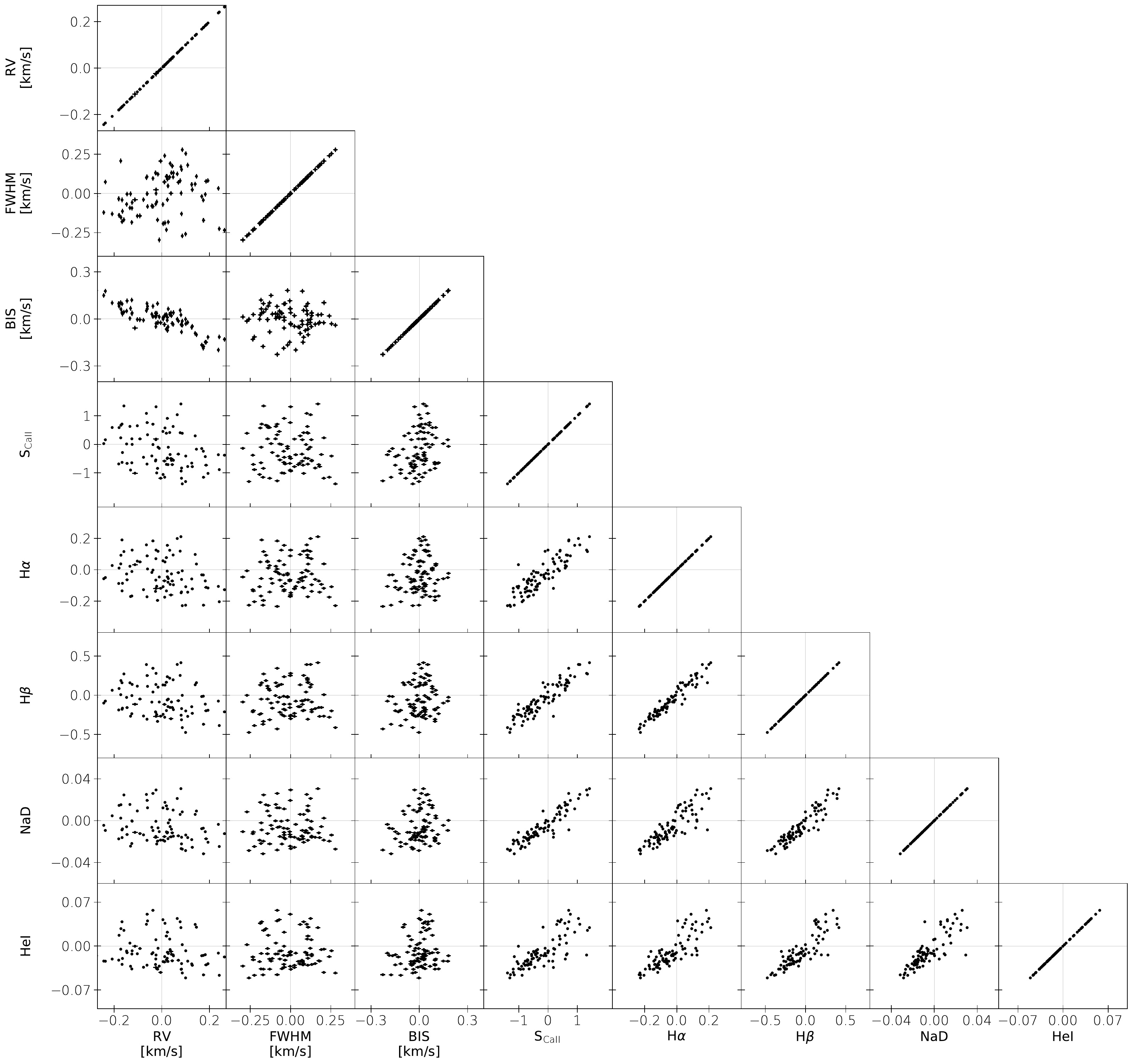}
    \caption{Correlation plots of the activity indicators analysed in Sec.~\ref{sec:section3}. The person correlation coefficients between each pair of time series are listed in Table~\ref{tab:correl_coeff}.}
    \label{fig:correl_plot}
\end{figure*}

Each pair of activity indicator computed for this study (i.e., RV, FWHM, BIS, and chromospheric emission indices based on Ca\,II H \& K, H$\alpha$, H$\beta$, Na\,I D1 \& D2 and He\,I D3 lines) is plotted in Fig.~\ref{fig:correl_plot} whereas the associated Pearson correlation coefficient are given in Table~\ref{tab:correl_coeff}.

\begin{table}
    \centering
    \caption{Pearson correlation coefficients between the different time series of activity indicators after removing the epochs affected by flares.}
    \label{tab:correl_coeff}
    \begin{tabular}{cccccccc}
        \hline
        & RV & FWHM & BIS & S$_{\rm{CaII}}$ & H$\alpha$ & H$\beta$ & Na\,I \\
        \hline
        FWHM & 0.124 &  -- &  -- &  -- &  -- &  -- &  -- \\
        BIS & -0.861 & -0.062 &  -- &  -- &  -- &  -- &  -- \\
        S$_{\rm{CaII}}$ & -0.336 & -0.066 & 0.263 &  -- &  -- &  -- &  -- \\
        H$\alpha$ & -0.245 & 0.101 & 0.169 & 0.846 &  -- &  -- &  -- \\
        H$\beta$ & -0.258 & 0.064 & 0.180 & 0.922 & 0.946 &  -- &  -- \\
        Na\, & -0.297 & -0.051 & 0.226 & 0.906 & 0.892 & 0.904 &  -- \\
        He\,I & -0.145 & 0.036 & 0.148 & 0.751 & 0.785 & 0.837 & 0.808 \\ 
        \hline
    \end{tabular}
\end{table}

\subsection{Periodograms}

\begin{figure*}
\begin{center}
        \includegraphics[width=0.6\textwidth]{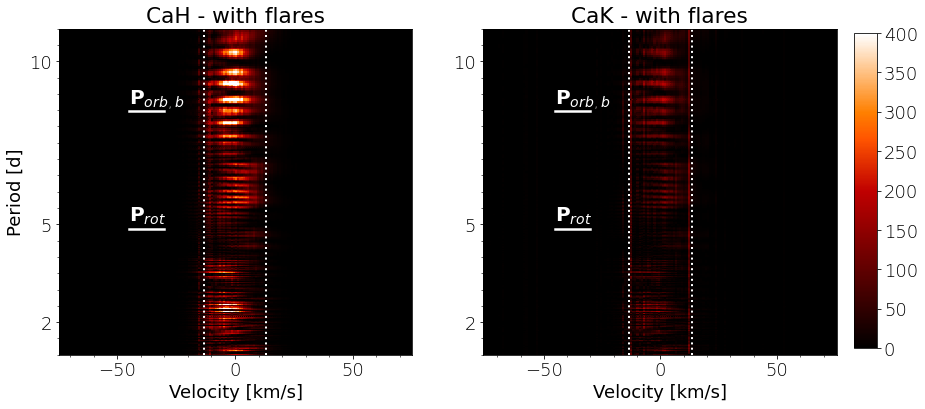}\
        \includegraphics[width=0.33\textwidth]{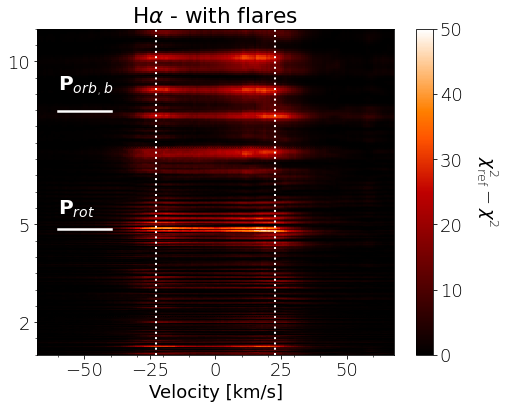}\\
        \includegraphics[width=0.6\textwidth]{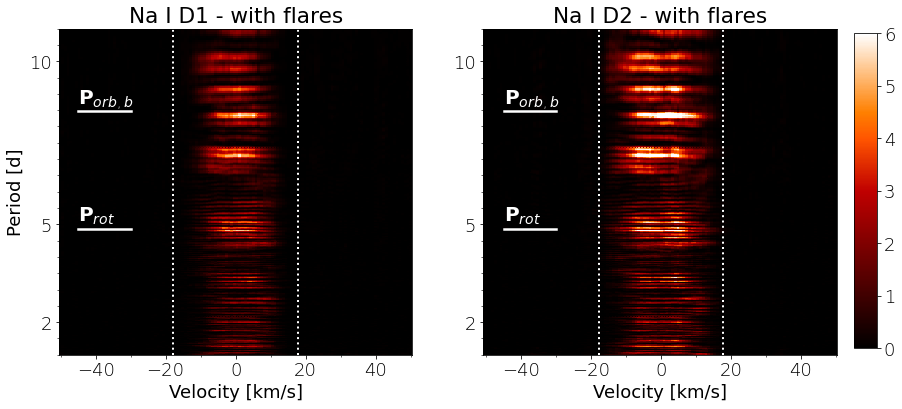}\ 
        \includegraphics[width=0.33\textwidth]{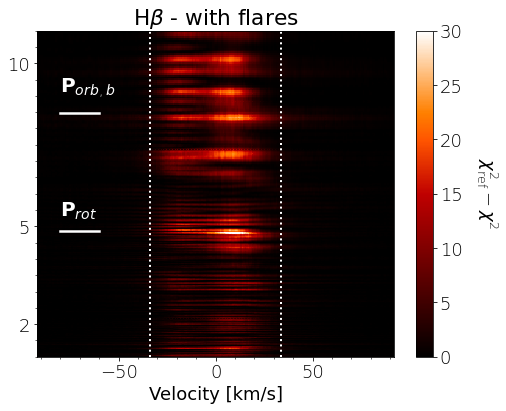} 
\caption{Bidimensional periodograms of the Ca\,II H \& K, H$\alpha$, Na\,I D1 \& D2 and H$\beta$ chromospheric lines, when all observations, including the ones affected by stellar flares, are included in the input data set. As in Fig.~\ref{fig:2D_period_he}, the two vertical lines delimit the line width and we indicate the orbital period of \aum\,b, \porbb, and the rotation period of the star, \pr, by the two horizontal lines. The color map depicts the power of the periodogram normalized to the most powerful peak.}
\label{fig:detail_2D_periodograms_fl}
\end{center}
\end{figure*}

\begin{figure*}
\begin{center}
        \includegraphics[width=0.6\textwidth]{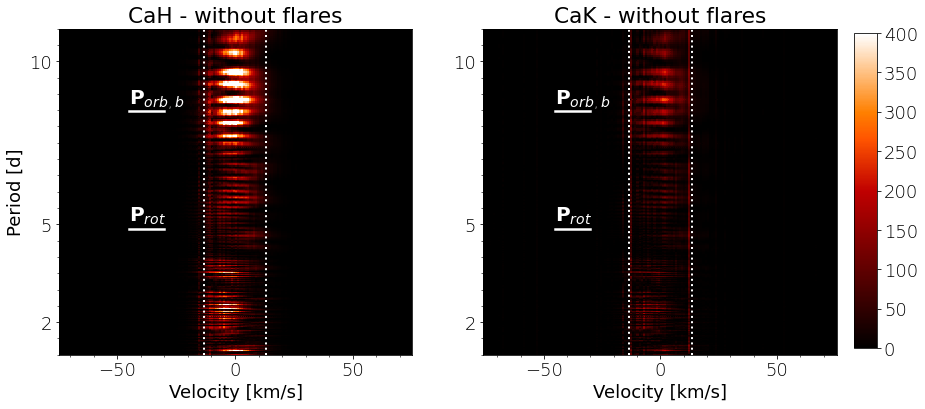}\
        \includegraphics[width=0.33\textwidth]{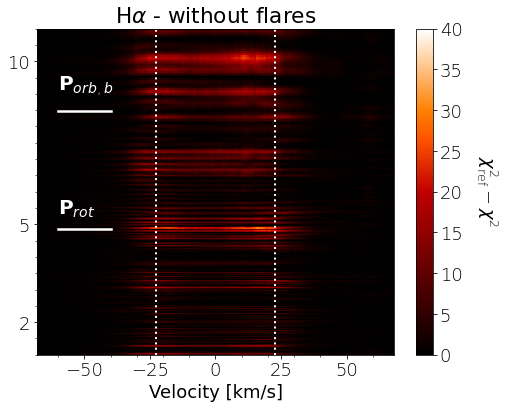}\\
        \includegraphics[width=0.6\textwidth]{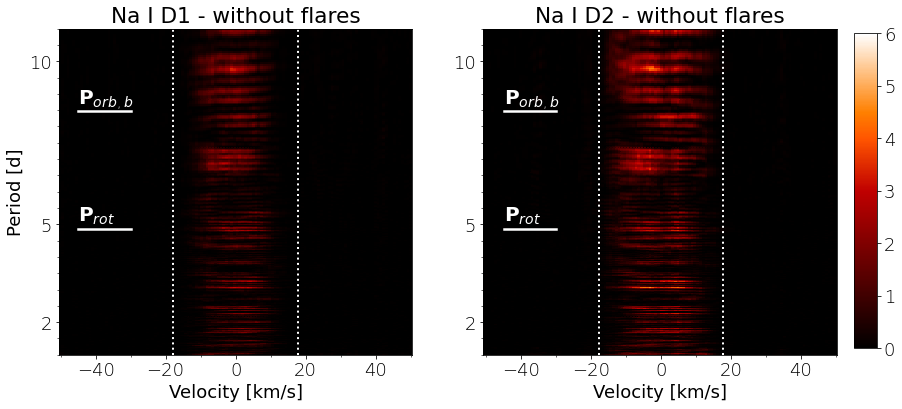}\ 
        \includegraphics[width=0.33\textwidth]{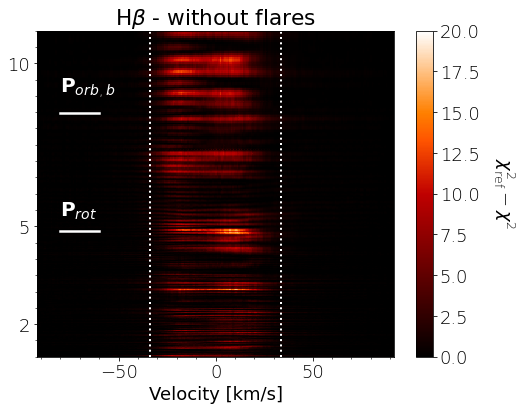} 
\caption{Same a Fig.~\ref{fig:detail_2D_periodograms_fl}, except that, this time, the observations affected by flares were excluded from the input data set beforehand.}
\label{fig:detail_2D_periodograms_nofl}
\end{center}
\end{figure*}

The Bidimensional periodograms the chromospheric emission lines except He I (shown in Fig.~\ref{fig:2D_period_he}) are shown in Fig.~\ref{fig:detail_2D_periodograms_fl} (including the observations affected by stellar flares in the input data set) and Fig.~\ref{fig:detail_2D_periodograms_nofl} (excluding the observations affected by stellar flares from the input data set).

\begin{figure}
    \centering
    \includegraphics[width=\linewidth]{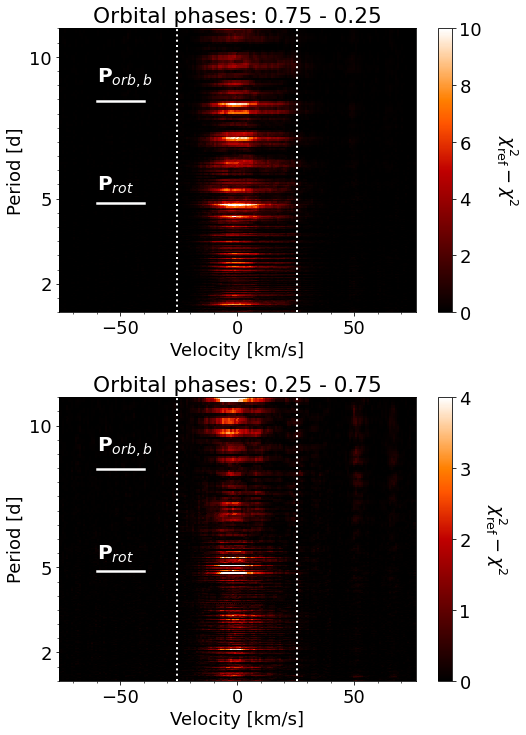}
    \caption{Bidimensional periodograms of the He\,I line obtained by only keeping the observations where \aum\,b is (i)~closer from the observer than the star (orbital phase between 0.75 and 0.25; \textit{top panel}), and (ii)~further from the observer than the star (orbital phase between 0.25 and 0.75; \textit{bottom panel}.}
    \label{fig:detail_2d_orb_phase}
\end{figure}


\bsp	
\label{lastpage}

\end{document}